\documentclass[11pt]{article}
\usepackage{graphicx}
\usepackage{amsmath}
\usepackage{amssymb}
\usepackage{caption2}
\begin{document}
\bibliographystyle{prsty}
\begin{center}
{\large {\bf \sc{  Analysis of  the $\Lambda_c(2860)$, $\Lambda_c(2880)$, $\Xi_c(3055)$  and $\Xi_c(3080)$  as D-wave baryon states with QCD sum rules }}} \\[2mm]
Zhi-Gang  Wang \footnote{E-mail: zgwang@aliyun.com. }   \\
 Department of Physics, North China Electric Power University,
Baoding 071003, P. R. China
\end{center}

\begin{abstract}
In this article, we tentatively assign the $\Lambda_c(2860)$, $\Lambda_c(2880)$, $\Xi_c(3055)$  and $\Xi_c(3080)$ to be the  D-wave baryon states with the spin-parity  $J^P={\frac{3}{2}}^+$, ${\frac{5}{2}}^+$, ${\frac{3}{2}}^+$  and ${\frac{5}{2}}^+$, respectively, and study their masses and pole residues  with the QCD sum rules in a systematic way  by constructing three-types interpolating currents with the quantum numbers $(L_\rho,L_\lambda)=(0,2)$, $(2,0)$ and $(1,1)$, respectively.
The present predictions favor  assigning the $\Lambda_c(2860)$, $\Lambda_c(2880)$, $\Xi_c(3055)$  and $\Xi_c(3080)$ to be the  D-wave baryon states with the quantum numbers $(L_\rho,L_\lambda)=(0,2)$ and $J^P={\frac{3}{2}}^+$, ${\frac{5}{2}}^+$, ${\frac{3}{2}}^+$  and ${\frac{5}{2}}^+$, respectively. While the predictions for the masses of  the $(L_\rho,L_\lambda)=(2,0)$ and $(1,1)$ D-wave $\Lambda_c$ and $\Xi_c$ states can be confronted to the experimental data in the future.
\end{abstract}



\section{Introduction}

Recently, the LHCb collaboration studied the mass spectrum of excited $\Lambda_c^+$  states that decay into $D^0 p$, and observed a new resonance $\Lambda_c(2860)^+$ near threshold \cite{LHCb2860}. The measured masses, widths and quantum numbers of the $\Lambda_c(2860)^+$, $\Lambda_c(2880)^+$   and $\Lambda_c(2940)^+$   states  are
\begin{flalign}
& \Lambda_c(2860) : M = 2856.1^{+2.0}_{-1.7}\pm 0.5 {}^{+1.1}_{-5.6} \mbox{ MeV}\, , \, \Gamma =67.6^{+10.1}_{-8.1} \pm 1.4{}^{+5.9}_{-20.0} \mbox{ MeV} \, , \,J^P={\frac{3}{2}}^+\, , \nonumber \\
& \Lambda_c(2880) : M =2881.75 \pm 0.29 \pm 0.07 {}^{+0.14}_{-0.20} \mbox{ MeV} \, , \, \Gamma = 5.43^{+0.77}_{-0.71}\pm 0.29 {}^{+0.75}_{-0.00} \mbox{ MeV}\, , \, J^P={\frac{5}{2}}^+ \, , \nonumber \\
& \Lambda_c(2940) : M = 2944.8^{+3.5}_{-2.5} \pm 0.4 {}^{+0.1}_{-4.6} \mbox{ MeV}\, , \, \Gamma = 27.7^{+8.2}_{-6.0}\pm 0.9 {}^{+5.2}_{-10.4} \mbox{ MeV} \, , \,J^P={\frac{3}{2}}^-\,  ,
\end{flalign}
but other assignments with the spins $J=\frac{1}{2}$ to $\frac{7}{2}$ are not excluded for the $\Lambda_c(2940)^+$ \cite{LHCb2860}.
The $\Lambda_c(2880)^+$ was first observed by the CLEO collaboration in the $\Lambda_c^+\pi^+\pi^-$ channel \cite{Artuso:2000xy}, confirmed by the BaBar collaboration in the $D^0p$ channel \cite{Aubert:2006sp} and the Belle collaboration in the $\Sigma_c(2455/2520)^{0,++}\pi^{+,-}$ channels \cite{Abe:2006rz}. The available experimental analysis indicates that the $\Lambda_c(2880)^+$ has the spin-parity  $J^P={\frac{5}{2}}^+$.
The theoretical predictions for the masses of the D-wave $\Lambda_c^+$ baryon states with $J^P={\frac{3}{2}}^+$ and ${\frac{5}{2}}^+$ are about $(2.85-2.90)\,\rm{GeV}$ \cite{Roberts2007,Ebert2011,Shah1602,Zhu-D-wave,Chen1609,Chen1701}. The $\Lambda_c(2860)^+$ and $\Lambda_c(2880)^+$ can be assigned to be the D-wave charmed  baryon states.

 Their strange cousins $\Xi_c(3055)^+$ and $\Xi_c(3080)^+$ were observed in the channel  $\Lambda_c^+K^-\pi^+$  by the Belle collaboration \cite{Chistov:2006zj} and in the channels $\Sigma_c(2455/2520)^{++}K^-$ by the BaBar collaboration \cite{Aubert:2007dt}. In 2016, the $\Xi_c(3055)^+$ and $\Xi_c(3055)^0$ were first observed by the Belle collaboration in the $D^+\Lambda$ and $D^0\Lambda$ channels, respectively \cite{Kato:2016hca}, the measured masses and widths were
\begin{flalign}
&\Xi_c(3055)^+: M=3055.8\pm0.4\pm0.2\mbox{ MeV}\, ,\, \Gamma=7.8\pm1.2\pm1.5\mbox{ MeV} \, ,\nonumber \\
&\Xi_c(3055)^0: M=3059.0\pm0.5\pm0.6\mbox{ MeV}\, ,\, \Gamma=6.4\pm 2.1\pm 1.1\mbox{ MeV} \, ,
\end{flalign}
furthermore, the Belle collaboration observed the first evidence for the $\Xi_c(3080)^+$ with the estimated  mass $3077.9\pm0.9\,\rm{MeV}$  and width  $3.0\pm0.7\pm0.4\,\rm{MeV}$.  The theoretical predictions of the masses of the D-wave $\Xi_c$ baryon states with $J^P={\frac{3}{2}}^+$ and ${\frac{5}{2}}^+$ are about $(3.05-3.10)\,\rm{GeV}$ \cite{Roberts2007,Ebert2011,Shah1602,Zhu-D-wave,Chen1609,Chen1701}, the $\Xi_c(3055)^+$, $\Xi_c(3055)^0$ and $\Xi_c(3080)^+$ can be assigned to be the D-wave charmed baryon states.

In this article, we tentatively assign the $\Lambda_c(2860)$, $\Lambda_c(2880)$, $\Xi_c(3055)$  and $\Xi_c(3080)$ to be the  D-wave charmed baryon states with the spin-parity $J^P={\frac{3}{2}}^+$, ${\frac{5}{2}}^+$, ${\frac{3}{2}}^+$  and ${\frac{5}{2}}^+$, respectively, and study their  masses and pole residues with the QCD sum rules in a systematic way. The QCD sum rules is a powerful theoretical approach in studying the ground state mass spectrum of the  heavy baryon states, and has given many successful descriptions \cite{Zhu-D-wave,WangNegativeP,WangNegativeCTP,Azizi-Negative,WangHbaryon,Zhu-P-wave,Wang-2625-2815}.

 We can construct the interpolating currents without introducing the relative P-wave to study the negative parity heavy, doubly-heavy and triply-heavy   baryon states \cite{WangNegativeP,WangNegativeCTP,Azizi-Negative}, or introducing the relative P-wave explicitly to study the negative parity heavy, doubly-heavy and triply-heavy   baryon states \cite{Zhu-P-wave,Wang-2625-2815}. For the D-wave heavy baryon states, it is better to introduce the relative D-wave explicitly to study them with the QCD sum rules \cite{Zhu-D-wave}. In Ref.\cite{Zhu-D-wave}, Chen et al study the mass spectrum  of the D-wave heavy baryon states with the QCD sum rules combined with the heavy quark effective theory in a systematic way. In this article, we  study the $\Lambda_c(2860)$, $\Lambda_c(2880)$, $\Xi_c(3055)$  and $\Xi_c(3080)$ as the D-wave heavy baryon states with the full QCD sum rules by introducing the relative D-wave explicitly in constructing the interpolating currents, which differ from the currents constructed in Ref.\cite{Zhu-D-wave}.

 The article is arranged as follows:  we derive the QCD sum rules for the masses and pole residues of  the D-wave ${\frac{3}{2}}^+$ and ${\frac{5}{2}}^+$ charmed  baryon states in Sect.2;  in Sect.3, we present the numerical results and discussions; and Sect.4 is reserved for our conclusion.

\section{QCD sum rules for  the D-wave ${\frac{3}{2}}^+$ and ${\frac{5}{2}}^+$ charmed  baryon states}
Firstly, we write down  the two-point correlation functions  $\Pi_{\alpha\beta}(p)$ and  $\Pi_{\alpha\beta\mu\nu}(p)$  in the QCD sum rules,
\begin{eqnarray}
\Pi_{\alpha\beta}(p)&=&i\int d^4x e^{ip \cdot x} \langle0|T\left\{J/\eta_{\alpha}(x)\bar{J}/\bar{\eta}_{\beta}(0)\right\}|0\rangle \, , \nonumber\\
\Pi_{\alpha\beta\mu\nu}(p)&=&i\int d^4x e^{ip \cdot x} \langle0|T\left\{J/\eta_{\alpha\beta}(x)\bar{J}/\bar{\eta}_{\mu\nu}(0)\right\}|0\rangle \, ,
\end{eqnarray}
where $J/\eta_\alpha(x)=J/\eta_\alpha^1(x),\,J/\eta_\alpha^2(x),\,J/\eta_\alpha^3(x)$, $J/\eta_{\alpha\beta}(x)=J/\eta_{\alpha\beta}^1(x),\,J/\eta_{\alpha\beta}^2(x),\,J/\eta_{\alpha\beta}^3(x)$,
\begin{eqnarray}
J^1_{\alpha}(x)&=&\varepsilon^{ijk} \left[ \partial^\mu \partial^\nu q^T_i(x) C\gamma_5 s_j(x)+\partial^\mu  q^T_i(x) C\gamma_5 \partial^\nu s_j(x)+\partial^\nu  q^T_i(x) C\gamma_5 \partial^\mu s_j(x)\right.\nonumber\\
&&\left.+ q^T_i(x) C\gamma_5 \partial^\mu \partial^\nu s_j(x)\right]\Gamma_{\mu\nu\alpha} c_k(x) \, ,\nonumber \\
J^2_{\alpha}(x)&=&\varepsilon^{ijk} \left[ \partial^\mu \partial^\nu q^T_i(x) C\gamma_5 s_j(x)-\partial^\mu  q^T_i(x) C\gamma_5 \partial^\nu s_j(x)-\partial^\nu  q^T_i(x) C\gamma_5 \partial^\mu s_j(x)\right.\nonumber\\
&&\left.+ q^T_i(x) C\gamma_5 \partial^\mu \partial^\nu s_j(x)\right]\Gamma_{\mu\nu\alpha} c_k(x) \, ,\nonumber \\
J^3_{\alpha}(x)&=&\varepsilon^{ijk} \left[ \partial^\mu \partial^\nu q^T_i(x) C\gamma_5 s_j(x)- q^T_i(x) C\gamma_5 \partial^\mu \partial^\nu s_j(x)\right]\Gamma_{\mu\nu\alpha} c_k(x) \, ,
\end{eqnarray}

\begin{eqnarray}
\eta^1_{\alpha}(x)&=&\varepsilon^{ijk} \left[ \partial^\mu \partial^\nu q^T_i(x) C\gamma_5 q^\prime_j(x)+\partial^\mu  q^T_i(x) C\gamma_5 \partial^\nu q^\prime_j(x)+\partial^\nu  q^T_i(x) C\gamma_5 \partial^\mu q^\prime_j(x)\right.\nonumber\\
&&\left.+ q^T_i(x) C\gamma_5 \partial^\mu \partial^\nu q^\prime_j(x)\right]\Gamma_{\mu\nu\alpha} c_k(x) \, ,\nonumber \\
\eta^2_{\alpha}(x)&=&\varepsilon^{ijk} \left[ \partial^\mu \partial^\nu q^T_i(x) C\gamma_5 q^\prime_j(x)-\partial^\mu  q^T_i(x) C\gamma_5 \partial^\nu q^\prime_j(x)-\partial^\nu  q^T_i(x) C\gamma_5 \partial^\mu q^\prime_j(x)\right.\nonumber\\
&&\left.+ q^T_i(x) C\gamma_5 \partial^\mu \partial^\nu q^\prime_j(x)\right]\Gamma_{\mu\nu\alpha} c_k(x) \, ,\nonumber \\
\eta^3_{\alpha}(x)&=&\varepsilon^{ijk} \left[ \partial^\mu \partial^\nu q^T_i(x) C\gamma_5 q^\prime_j(x)- q^T_i(x) C\gamma_5 \partial^\mu \partial^\nu q^\prime_j(x)\right]\Gamma_{\mu\nu\alpha} c_k(x) \, ,
\end{eqnarray}

\begin{eqnarray}
J^1_{\alpha\beta}(x)&=&\varepsilon^{ijk} \left[ \partial^\mu \partial^\nu q^T_i(x) C\gamma_5 s_j(x)+\partial^\mu  q^T_i(x) C\gamma_5 \partial^\nu s_j(x)+\partial^\nu  q^T_i(x) C\gamma_5 \partial^\mu s_j(x)\right.\nonumber\\
&&\left.+ q^T_i(x) C\gamma_5 \partial^\mu \partial^\nu s_j(x)\right]\Gamma_{\mu\nu\alpha\beta} c_k(x) \, ,\nonumber \\
J^2_{\alpha\beta}(x)&=&\varepsilon^{ijk} \left[ \partial^\mu \partial^\nu q^T_i(x) C\gamma_5 s_j(x)-\partial^\mu  q^T_i(x) C\gamma_5 \partial^\nu s_j(x)-\partial^\nu  q^T_i(x) C\gamma_5 \partial^\mu s_j(x)\right.\nonumber\\
&&\left.+ q^T_i(x) C\gamma_5 \partial^\mu \partial^\nu s_j(x)\right]\Gamma_{\mu\nu\alpha\beta} c_k(x) \, ,\nonumber \\
J^3_{\alpha\beta}(x)&=&\varepsilon^{ijk} \left[ \partial^\mu \partial^\nu q^T_i(x) C\gamma_5 s_j(x)- q^T_i(x) C\gamma_5 \partial^\mu \partial^\nu s_j(x)\right]\Gamma_{\mu\nu\alpha\beta} c_k(x) \, ,
\end{eqnarray}

\begin{eqnarray}
\eta^1_{\alpha\beta}(x)&=&\varepsilon^{ijk} \left[ \partial^\mu \partial^\nu q^T_i(x) C\gamma_5 q^\prime_j(x)+\partial^\mu  q^T_i(x) C\gamma_5 \partial^\nu q^\prime_j(x)+\partial^\nu  q^T_i(x) C\gamma_5 \partial^\mu q^\prime_j(x)\right.\nonumber\\
&&\left.+ q^T_i(x) C\gamma_5 \partial^\mu \partial^\nu q^\prime_j(x)\right]\Gamma_{\mu\nu\alpha\beta} c_k(x) \, ,\nonumber \\
\eta^2_{\alpha\beta}(x)&=&\varepsilon^{ijk} \left[ \partial^\mu \partial^\nu q^T_i(x) C\gamma_5 q^\prime_j(x)-\partial^\mu  q^T_i(x) C\gamma_5 \partial^\nu q^\prime_j(x)-\partial^\nu  q^T_i(x) C\gamma_5 \partial^\mu q^\prime_j(x)\right.\nonumber\\
&&\left.+ q^T_i(x) C\gamma_5 \partial^\mu \partial^\nu q^\prime_j(x)\right]\Gamma_{\mu\nu\alpha\beta} c_k(x) \, ,\nonumber \\
\eta^3_{\alpha\beta}(x)&=&\varepsilon^{ijk} \left[ \partial^\mu \partial^\nu q^T_i(x) C\gamma_5 q^\prime_j(x)- q^T_i(x) C\gamma_5 \partial^\mu \partial^\nu q^\prime_j(x)\right]\Gamma_{\mu\nu\alpha\beta} c_k(x) \, ,
\end{eqnarray}
with
\begin{eqnarray}
\Gamma_{\mu\nu\alpha}&=&\left(g_{\mu\alpha}g_{\nu\beta} +g_{\mu\beta}g_{\nu\alpha}-\frac{1}{2}g_{\mu\nu}g_{\alpha\beta}\right)\gamma^\beta\gamma_5 \, , \nonumber\\
\Gamma_{\mu\nu\alpha\beta}&=&g_{\mu\alpha}g_{\nu\beta} +g_{\mu\beta}g_{\nu\alpha}-\frac{1}{6}g_{\mu\nu}g_{\alpha\beta}-\frac{1}{4}g_{\mu\alpha}\gamma_\nu\gamma_\beta-\frac{1}{4}g_{\mu\beta}\gamma_\nu\gamma_\alpha
-\frac{1}{4}g_{\nu\alpha}\gamma_\mu\gamma_\beta-\frac{1}{4}g_{\nu\beta}\gamma_\mu\gamma_\alpha \nonumber\\
&&+\frac{1}{24}\gamma_\mu\gamma_\alpha\gamma_\nu\gamma_\beta+\frac{1}{24}\gamma_\mu\gamma_\beta\gamma_\nu\gamma_\alpha
+\frac{1}{24}\gamma_\nu\gamma_\alpha\gamma_\mu\gamma_\beta+\frac{1}{24}\gamma_\nu\gamma_\beta\gamma_\mu\gamma_\alpha\, ,
\end{eqnarray}
$q,\, q^\prime=u,\,d$, the $i$, $j$, $k$ are color indices, the $C$ is the charge conjugation matrix.  The currents satisfy the relations $\gamma^\alpha J^i_\alpha(x)=\gamma^\alpha \eta^i_\alpha(x)=0$,  $\gamma^\alpha J^i_{\alpha\beta}(x)=\gamma^\alpha \eta^i_{\alpha\beta}(x)=0$, $\gamma^\beta J^i_{\alpha\beta}(x)=\gamma^\beta \eta^i_{\alpha\beta}(x)=0$, where $i=1,2,3$.  We choose the currents $J/\eta^i_\alpha(x)$ and $J/\eta^i_{\alpha\beta}(x)$ to interpolate the $J^P={\frac{3}{2}}^+$ and ${\frac{5}{2}}^+$ charmed  baryon states,  respectively.  In this article, we tentatively assign the $\Lambda_c(2860)$, $\Lambda_c(2880)$, $\Xi_c(3055)$  and $\Xi_c(3080)$ to be the  D-wave charmed baryon states with the spin-parity $J^P={\frac{3}{2}}^+$, ${\frac{5}{2}}^+$, ${\frac{3}{2}}^+$  and ${\frac{5}{2}}^+$, respectively, the currents $J_\alpha(x)$, $J_{\alpha\beta}(x)$, $\eta_\alpha(x)$ and $\eta_{\alpha\beta}(x)$ may couple potentially to the     $\Lambda_c(2860)$, $\Lambda_c(2880)$, $\Xi_c(3055)$  and $\Xi_c(3080)$, respectively.

Now we take a short digression to  illustrate how to construct the currents.
The attractive interaction of one-gluon exchange  favors  formation of
the diquarks in  color antitriplet $\overline{3}_{ c}$ \cite{One-gluon}.
The color antitriplet  diquarks $\varepsilon^{ijk} q^{T}_j C\Gamma q^{\prime}_k$
  have  five  structures  in Dirac spinor space, where $C\Gamma=C\gamma_5$, $C$, $C\gamma_\mu \gamma_5$,  $C\gamma_\mu $ and $C\sigma_{\mu\nu}$ for the scalar, pseudoscalar, vector, axialvector  and  tensor diquarks, respectively.  The structures
$C\gamma_\mu $ and $C\sigma_{\mu\nu}$ are symmetric, while the structures
$C\gamma_5$, $C$ and $C\gamma_\mu \gamma_5$ are antisymmetric.
The calculations based on the QCD sum rules  indicate that  the favored configurations are the $C\gamma_5$ and $C\gamma_\mu$ diquark states, while the most favored  configurations are the $C\gamma_5$   diquark states \cite{WangLDiquark}.

We usually construct the heavy baryon states according to the light-diquark-heavy-quark model. In the diquark-quark models, the angular momentum between the two light quarks is denoted by $L_\rho$, while the angular momentum between the  light diquark and the heavy quark is denoted by $L_\lambda$. If the two light quarks in the  diquark are in relative S-wave or $L_\rho=0$, then the baryons with the $J^P=0^+$ and $1^+$ diquarks (the ground state diquarks) are called $\Lambda$-type and $\Sigma$-type baryons, respectively \cite{Korner-PPNP}.  We can denote the $C\gamma_5$ and $C\gamma_\mu$  diquarks as $J_d^P=0_d^+$ and $1_d^+$, respectively,   the relative P-wave and D-wave as $J_{\rho/\lambda}^P=L_{\rho/\lambda}^P=1_{\rho/\lambda}^-$ and $2_{\rho/\lambda}^+$, respectively, the $c$-quark as $J_c^P={\frac{1}{2}}_c^+$, then we construct the D-wave baryon states according to the routines,
\begin{eqnarray}
0_d^+\otimes 2_{\rho/\lambda}^+ \otimes {\frac{1}{2}}_c^+ &= & \underline{2^+} \otimes {\frac{1}{2}}_c^+=  \underline{{\frac{3}{2}}^+\oplus {\frac{5}{2}}^+} \, , \\
0_d^+\otimes 1_\rho^- \otimes 1_\lambda^-  \otimes {\frac{1}{2}}_c^+ &= & \left[0^+\oplus1^+\oplus \underline{2^+} \right]   \otimes {\frac{1}{2}}_c^+= {\frac{1}{2}}^+\oplus\left[{\frac{1}{2}}^+\oplus{\frac{3}{2}}^+\right]\oplus\underline{\left[{\frac{3}{2}}^+\oplus {\frac{5}{2}}^+\right]} \, , \\
1_d^+\otimes 2_{\rho/\lambda}^+ \otimes {\frac{1}{2}}_c^+ &= & \left[1^+\oplus 2^+\oplus 3^+\right]\otimes {\frac{1}{2}}_c^+ =\left[{\frac{1}{2}}^+\oplus{\frac{3}{2}}^+\right]\oplus\left[{\frac{3}{2}}^+\oplus {\frac{5}{2}}^+\right]\oplus\left[{\frac{5}{2}}^+\oplus {\frac{7}{2}}^+\right] \, , \nonumber\\
1_d^+\otimes 1_\rho^- \otimes 1_\lambda^-  \otimes {\frac{1}{2}}_c^+ &= &  \left[0^-\oplus1^-\oplus 2^- \right] \otimes 1_\lambda^- \otimes  {\frac{1}{2}}_c^+\nonumber\\
&=& \left[1^+\oplus \left[0^+\oplus1^+\oplus2^+\right]\oplus \left[1^+\oplus2^+\oplus3^+ \right]\right]\otimes {\frac{1}{2}}_c^+\nonumber\\
&=&\left[{\frac{1}{2}}^+\oplus{\frac{3}{2}}^+\right]\oplus{\frac{1}{2}}^+\oplus\left[{\frac{1}{2}}^+\oplus{\frac{3}{2}}^+\right]\oplus\left[{\frac{3}{2}}^+\oplus {\frac{5}{2}}^+\right]\oplus\left[{\frac{1}{2}}^+\oplus{\frac{3}{2}}^+\right]\nonumber\\
&&\oplus\left[{\frac{3}{2}}^+\oplus {\frac{5}{2}}^+\right]\oplus\left[{\frac{5}{2}}^+\oplus {\frac{7}{2}}^+\right]\, .
\end{eqnarray}
It is difficult or impossible to construct  currents to interpolate all the D-wave baryon states with $J^P={\frac{1}{2}}^+$, ${\frac{3}{2}}^+$, ${\frac{5}{2}}^+$ and ${\frac{7}{2}}^+$ in a systematic way.
In this article, we study the underlined D-wave baryon states with $J^P={\frac{3}{2}}^+$ and ${\frac{5}{2}}^+$ in details based on the most favored  configurations $C\gamma_5$   \cite{WangLDiquark}. Experimentally, the measured  quantum numbers of the $\Lambda_c(2860)^+$ and $\Lambda_c(2880)^+$   are $J^P={\frac{3}{2}}^+$ and ${\frac{5}{2}}^+$ respectively from the LHCb collaboration \cite{LHCb2860}, while the masses of the $\Xi_c(3055)^+$, $\Xi_c(3055)^0$ and $\Xi_c(3080)^+$ are consistent with  the theoretical predictions of   the D-wave $\Xi_c$ baryon states with $J^P={\frac{3}{2}}^+$ and ${\frac{5}{2}}^+$ \cite{Roberts2007,Ebert2011,Shah1602,Zhu-D-wave,Chen1609,Chen1701}.

We  can choose either the partial  derivative $\partial_\mu$ or the covariant derivative $D_\mu$ to construct  the interpolating  currents. The currents with the covariant derivative $D_\mu$ are gauge invariant, but blur the physical interpretation of the $\stackrel{\leftrightarrow}{D}_\mu=\stackrel{\rightarrow}{\partial}_\mu-ig_sG_\mu-\stackrel{\leftarrow}{\partial}_\mu-ig_sG_\mu $ being the angular momentum. The currents with the partial derivative $\partial_\mu$ are not gauge invariant, but manifests the physical interpretation of the $\stackrel{\leftrightarrow}{\partial}_\mu=\stackrel{\rightarrow}{\partial}_\mu-\stackrel{\leftarrow}{\partial}_\mu$ being the angular momentum. In Ref.\cite{WangDi}, we study the masses and decay constants of the heavy tensor mesons  $D_2^*(2460)$, $D_{s2}^*(2573)$, $B_2^*(5747)$ and $B_{s2}^*(5840)$  with the QCD sum rules. In calculations, we observe that  the predictions based on the currents with the partial derivative and covariant derivative differ from each other about $1\%$, if the same parameters are chosen. If we refit the Borel parameters and threshold parameters, the  differences   about $1\%$ can be reduced remarkably, so the currents with the partial derivative work well. In this article, we choose the partial  derivative $\partial_\mu$ to construct the interpolating currents. Furthermore, from the Table 1 in Section
3, we can see that the dominant contributions come from the perturbative terms, so neglecting the contributions originate from the gluons in the covariant derivative $D_\mu$ cannot change the conclusion.

  For $L_\rho=1$ and $L_{\lambda}=0$, the light diquark state with $J^P=1^-$ can be written as
\begin{eqnarray}
\varepsilon^{ijk} \left[ \partial^\nu  q^T_i(x) C\gamma_5 q^\prime_j(x) -  q^T_i(x) C\gamma_5 \partial^\nu q^\prime_j(x)\right]\, ,
\end{eqnarray}
then we introduce an additional P-wave between the two quarks $q$ and $q^\prime$, and obtain the light diquark state with $L_\rho=2$, $L_{\lambda}=0$ and $J^P=2^+$,
\begin{eqnarray}
&&\varepsilon^{ijk} \left\{ \left[\partial^\mu\partial^\nu q^T_i(x) C\gamma_5 q^\prime_j(x)-\partial^\nu q^T_i(x) C\gamma_5 \partial^\mu q^\prime_j(x) \right]-  \left[\partial^\mu q^T_i(x) C\gamma_5 \partial^\nu q^\prime_j(x)\right.\right.\nonumber\\
&&\left.\left.-q^T_i(x) C\gamma_5\partial^\mu \partial^\nu q^\prime_j(x)\right]\right\} \, .
\end{eqnarray}
In the heavy quark limit, the $c$-quark is static, the $\stackrel{\leftrightarrow}{\partial}_\mu$ is reduced to $\stackrel{\leftarrow}{\partial}_\mu$ when operating  on the $c$-quark field.
For $L_\rho=0$ and $L_{\lambda}=2$, the light diquark state with $J^P=2^+$ can be written as
\begin{eqnarray}
\partial^\mu \partial^\nu\left[\varepsilon^{ijk}q^T_i(x) C\gamma_5 q^\prime_j(x)\right] &=&\varepsilon^{ijk} \left[\partial^\mu\partial^\nu q^T_i(x) C\gamma_5 q^\prime_j(x)+\partial^\mu q^T_i(x) C\gamma_5 \partial^\nu q^\prime_j(x)\right.\nonumber\\
&&\left.+\partial^\nu q^T_i(x) C\gamma_5 \partial^\mu q^\prime_j(x)+q^T_i(x) C\gamma_5\partial^\mu \partial^\nu q^\prime_j(x)\right] \, .
\end{eqnarray}
For $L_\rho=1$ and $L_{\lambda}=1$, the light diquark state with $J^P=2^+$ can be written as
\begin{eqnarray}
&&\partial^\mu\varepsilon^{ijk} \left[ \partial^\nu  q^T_i(x) C\gamma_5 q^\prime_j(x) -  q^T_i(x) C\gamma_5 \partial^\nu q^\prime_j(x)\right]\nonumber\\
&&=\varepsilon^{ijk} \left[ \partial^\mu\partial^\nu  q^T_i(x) C\gamma_5 q^\prime_j(x)+\partial^\nu  q^T_i(x) C\gamma_5 \partial^\mu q^\prime_j(x) -  \partial^\mu q^T_i(x) C\gamma_5 \partial^\nu q^\prime_j(x)\right.\nonumber\\
&&\left. -q^T_i(x) C\gamma_5 \partial^\mu\partial^\nu q^\prime_j(x) \right]\, .
\end{eqnarray}
We symmetrize the Lorentz indexes $\mu$ and $\nu$, and obtain the light diquark state with $L_\rho=1$ and $L_{\lambda}=1$ in a more simple form,
   \begin{eqnarray}
\varepsilon^{ijk} \left[ \partial^\mu\partial^\nu  q^T_i(x) C\gamma_5 q^\prime_j(x) -q^T_i(x) C\gamma_5 \partial^\mu\partial^\nu q^\prime_j(x) \right]\, .
\end{eqnarray}
The light diquark states with  $J^P=2^+$ then combine with the $c$-quark to form   $J^P={\frac{3}{2}}^+$ or ${\frac{5}{2}}^+$ baryon states, see Eqs.(9-10).

The interpolating currents can be classified by
\begin{eqnarray}
(L_{\rho},L_{\lambda})=(0,2)\,\,&{\rm for} & \,\,J/\eta^1_{\alpha}(x)\, ,\,\,J/\eta^1_{\alpha\beta}(x) \, , \nonumber\\
(L_{\rho},L_{\lambda})=(2,0)\,\,&{\rm for} & \,\,J/\eta^2_{\alpha}(x)\, ,\,\,J/\eta^2_{\alpha\beta}(x) \, , \nonumber\\
(L_{\rho},L_{\lambda})=(1,1)\,\,&{\rm for} & \,\,J/\eta^3_{\alpha}(x)\, ,\,\,J/\eta^3_{\alpha\beta}(x) \, .
\end{eqnarray}

 The currents  $J/\eta_\alpha(0)$ and $J/\eta_{\alpha\beta}(0)$ couple potentially to the   ${\frac{3}{2}}^\pm$ and  ${\frac{5}{2}}^\pm$  charmed baryon
 states   $B_{\frac{3}{2}}^{\pm}$ and   $B_{\frac{5}{2}}^{\pm}$, respectively \cite{WangHbaryon,Oka96,WangPc}, which are supposed to be the excited $\Lambda_c$ or $\Xi_c$ states,
\begin{eqnarray}
\langle 0| J/\eta_{\alpha} (0)|B_{\frac{3}{2}}^{+}(p)\rangle &=&\lambda^{+}_{\frac{3}{2}} U^{+}_\alpha(p,s) \, ,  \nonumber\\
\langle 0| J/\eta_{\alpha\beta} (0)|B_{\frac{5}{2}}^{+}(p)\rangle &=&\lambda^{+}_{\frac{5}{2}} U^{+}_{\alpha\beta}(p,s) \, ,\\
\langle 0| J/\eta_{\alpha} (0)|B_{\frac{3}{2}}^{-}(p)\rangle &=&\lambda^{-}_{\frac{3}{2}}i\gamma_5 U^{-}_{\alpha}(p,s) \, , \nonumber\\
\langle 0| J/\eta_{\alpha\beta} (0)|B_{\frac{5}{2}}^{-}(p)\rangle &=&\lambda^{-}_{\frac{5}{2}}i\gamma_5 U^{-}_{\alpha\beta}(p,s) \, ,
\end{eqnarray}
where the $\lambda^{\pm}_{\frac{3}{2}}$ and $\lambda^{\pm}_{\frac{5}{2}}$ are the pole residues or the current-baryon coupling constants, the  spinors $U^{\pm}_\alpha(p,s)$ and $U^{\pm}_{\alpha\beta}(p,s)$ satisfy the Rarita-Schwinger equations $(\not\!\!p-M_{\pm})U^{\pm}_\alpha(p,s)=0$ and $(\not\!\!p-M_{\pm})U^{\pm}_{\alpha\beta}(p,s)=0$,  and the relations $\gamma^\alpha U^{\pm}_\alpha(p,s)=0$,
$p^\alpha U^{\pm}_\alpha(p,s)=0$, $\gamma^\alpha U^{\pm}_{\alpha\beta}(p,s)=0$,
$p^\alpha U^{\pm}_{\alpha\beta}(p,s)=0$, $ U^{\pm}_{\alpha\beta}(p,s)= U^{\pm}_{\beta\alpha}(p,s)$, which are consistent with relations $\gamma^\alpha J^i_\alpha(0)=\gamma^\alpha \eta^i_\alpha(0)=0$,  $\gamma^\alpha J^i_{\alpha\beta}(0)=\gamma^\alpha \eta^i_{\alpha\beta}(0)=0$, $\gamma^\beta J^i_{\alpha\beta}(0)=\gamma^\beta \eta^i_{\alpha\beta}(0)=0$.

At the hadron side,  we  insert  a complete set  of intermediate charmed baryon  states with the
same quantum numbers as the current operators  $J/\eta_\alpha(x)$,
$i\gamma_5 J/\eta_\alpha(x)$, $J/\eta_{\alpha\beta}(x)$ and
$i\gamma_5 J/\eta_{\alpha\beta}(x)$ into the correlation functions
$\Pi_{\alpha\beta}(p)$ and $\Pi_{\alpha\beta\mu\nu}(p)$ to obtain the hadronic representation
\cite{SVZ79,PRT85}. We isolate the pole terms of the lowest
 charmed  baryon states with  positive parity and negative parity,    and obtain the  results:
\begin{eqnarray}
\Pi_{\alpha\beta}(p) & = & {\lambda^{+}_{\frac{3}{2}}}^2  {\!\not\!{p}+ M_{+} \over M_{+}^{2}-p^{2}  } \left(- g_{\alpha\beta}+\frac{\gamma_\alpha\gamma_\beta}{3}+\frac{2p_\alpha p_\beta}{3p^2}-\frac{p_\alpha\gamma_\beta-p_\beta \gamma_\alpha}{3\sqrt{p^2}}
\right)\nonumber\\
&&+  {\lambda^{-}_{\frac{3}{2}}}^2  {\!\not\!{p}- M_{-} \over M_{-}^{2}-p^{2}  } \left(- g_{\alpha\beta}+\frac{\gamma_\alpha\gamma_\beta}{3}+\frac{2p_\alpha p_\beta}{3p^2}-\frac{p_\alpha\gamma_\beta-p_\beta \gamma_\alpha}{3\sqrt{p^2}}
\right) +\cdots  \, ,\nonumber\\
&=&\Pi_{\frac{3}{2}}(p^2)\,\left(- g_{\alpha\beta}\right)+\cdots\, ,
\end{eqnarray}
\begin{eqnarray}
\Pi_{\alpha\beta\mu\nu}(p) & = & {\lambda^{+}_{\frac{5}{2}}}^2  {\!\not\!{p}+ M_{+} \over M_{+}^{2}-p^{2}  } \left[\frac{ \widetilde{g}_{\mu\alpha}\widetilde{g}_{\nu\beta}+\widetilde{g}_{\mu\beta}\widetilde{g}_{\nu\alpha}}{2}-\frac{\widetilde{g}_{\mu\nu}\widetilde{g}_{\alpha\beta}}{5}
-\frac{1}{10}\left( \gamma_{\alpha}\gamma_{\mu}+\frac{\gamma_{\alpha}p_{\mu}-\gamma_{\mu}p_{\alpha}}{\sqrt{p^2}}-\frac{p_{\alpha}p_{\mu}}{p^2}\right)\widetilde{g}_{\nu\beta}\right.\nonumber\\
&&\left.-\frac{1}{10}\left( \gamma_{\alpha}\gamma_{\nu}+\frac{\gamma_{\alpha}p_{\nu}-\gamma_{\nu}p_{\alpha}}{\sqrt{p^2}}-\frac{p_{\alpha}p_{\nu}}{p^2}\right)\widetilde{g}_{\mu\beta}
+\cdots\right]\nonumber\\
&&+   {\lambda^{-}_{\frac{5}{2}}}^2  {\!\not\!{p}- M_{-} \over M_{-}^{2}-p^{2}  }  \left[\frac{ \widetilde{g}_{\mu\alpha}\widetilde{g}_{\nu\beta}+\widetilde{g}_{\mu\beta}\widetilde{g}_{\nu\alpha}}{2}-\frac{\widetilde{g}_{\mu\nu}\widetilde{g}_{\alpha\beta}}{5}
-\frac{1}{10}\left( \gamma_{\alpha}\gamma_{\mu}+\frac{\gamma_{\alpha}p_{\mu}-\gamma_{\mu}p_{\alpha}}{\sqrt{p^2}}-\frac{p_{\alpha}p_{\mu}}{p^2}\right)\widetilde{g}_{\nu\beta}\right.\nonumber\\
&&\left.-\frac{1}{10}\left( \gamma_{\alpha}\gamma_{\nu}+\frac{\gamma_{\alpha}p_{\nu}-\gamma_{\nu}p_{\alpha}}{\sqrt{p^2}}-\frac{p_{\alpha}p_{\nu}}{p^2}\right)\widetilde{g}_{\mu\beta}
+\cdots\right]   +\cdots \, , \nonumber\\
&=&\Pi_{\frac{5}{2}}(p^2)\,\frac{ g_{\mu\alpha}g_{\nu\beta}+g_{\mu\beta}g_{\nu\alpha}}{2}+\cdots \, ,
\end{eqnarray}
where $\widetilde{g}_{\mu\nu}=g_{\mu\nu}-\frac{p_{\mu}p_{\nu}}{p^2}$.
The currents $J/\eta_\alpha(0)$, $J/\eta_{\alpha\beta}(0)$ also have non-vanishing couplings with  the spin $J=\frac{1}{2}$ and $J=\frac{1}{2}, \frac{3}{2}$ charmed baryon states, respectively,  we choose the tensor structures $g_{\alpha\beta}$ and $g_{\mu\alpha}g_{\nu\beta}+g_{\mu\beta}g_{\nu\alpha}$ for analysis, the baryon states with the spin $\frac{1}{2}$ and $\frac{3}{2}$  have no contaminations \cite{WangPc}.

In calculations, we have used two  summations over the polarizations $s$ in the spinors $U^{\pm}_\alpha(p,s)$ and $U^{\pm}_{\alpha\beta}(p,s)$ \cite{HuangShiZhong},
\begin{eqnarray}
\sum_s U_\alpha \overline{U}_\beta&=&\left(\!\not\!{p}+M_{\pm}\right)\left( -g_{\alpha\beta}+\frac{\gamma_\alpha\gamma_\beta}{3}+\frac{2p_\alpha p_\beta}{3p^2}-\frac{p_\alpha\gamma_\beta-p_\beta \gamma_\alpha}{3\sqrt{p^2}} \right) \,  ,  \\
\sum_s U_{\mu\nu}\overline {U}_{\alpha\beta}&=&\left(\!\not\!{p}+M_{\pm}\right)\left\{\frac{\widetilde{g}_{\mu\alpha}\widetilde{g}_{\nu\beta}+\widetilde{g}_{\mu\beta}\widetilde{g}_{\nu\alpha}}{2} -\frac{\widetilde{g}_{\mu\nu}\widetilde{g}_{\alpha\beta}}{5}-\frac{1}{10}\left( \gamma_{\mu}\gamma_{\alpha}+\frac{\gamma_{\mu}p_{\alpha}-\gamma_{\alpha}p_{\mu}}{\sqrt{p^2}}-\frac{p_{\mu}p_{\alpha}}{p^2}\right)\widetilde{g}_{\nu\beta}\right. \nonumber\\
&&-\frac{1}{10}\left( \gamma_{\nu}\gamma_{\alpha}+\frac{\gamma_{\nu}p_{\alpha}-\gamma_{\alpha}p_{\nu}}{\sqrt{p^2}}-\frac{p_{\nu}p_{\alpha}}{p^2}\right)\widetilde{g}_{\mu\beta}
-\frac{1}{10}\left( \gamma_{\mu}\gamma_{\beta}+\frac{\gamma_{\mu}p_{\beta}-\gamma_{\beta}p_{\mu}}{\sqrt{p^2}}-\frac{p_{\mu}p_{\beta}}{p^2}\right)\widetilde{g}_{\nu\alpha}\nonumber\\
&&\left.-\frac{1}{10}\left( \gamma_{\nu}\gamma_{\beta}+\frac{\gamma_{\nu}p_{\beta}-\gamma_{\beta}p_{\nu}}{\sqrt{p^2}}-\frac{p_{\nu}p_{\beta}}{p^2}\right)\widetilde{g}_{\mu\alpha} \right\} \, ,
\end{eqnarray}
and $p^2=M^2_{\pm}$ on  mass-shell.

We obtain the hadronic spectral densities at the hadron  side through dispersion relation,
\begin{eqnarray}
\frac{{\rm Im}\Pi_{j}(s)}{\pi}&=&\!\not\!{p} \left[{\lambda^{+}_{j}}^2 \delta\left(s-M_{+}^2\right)+{\lambda^{-}_{j}}^2 \delta\left(s-M_{-}^2\right)\right] +\left[M_{+}{\lambda^{+}_{j}}^2 \delta\left(s-M_{+}^2\right)-M_{-}{\lambda^{-}_{j}}^2 \delta\left(s-M_{-}^2\right)\right]\, , \nonumber\\
&=&\!\not\!{p}\, \rho^1_{j,H}(s)+\rho^0_{j,H}(s) \, ,
\end{eqnarray}
where $j=\frac{3}{2}$, $\frac{5}{2}$, the subscript $H$ denotes  the hadron side,
then we introduce the exponential function $\exp\left(-\frac{s}{T^2}\right)$ to depress the continuum state contributions  to obtain the QCD sum rules at the hadron side,
\begin{eqnarray}
\int_{m_c^2}^{s_0}ds \left[\sqrt{s}\rho^1_{j,H}(s)+\rho^0_{j,H}(s)\right]\exp\left( -\frac{s}{T^2}\right)
&=&2M_{+}{\lambda^{+}_{j}}^2\exp\left( -\frac{M_{+}^2}{T^2}\right) \, ,
\end{eqnarray}
where the $s_0$ are the continuum thresholds and the $T^2$ are the Borel parameters \cite{WangPc}. From Eq.(25), we can see that the ${\frac{3}{2}}^{-}$ and ${\frac{5}{2}}^{-}$ charmed baryon states have no contaminations according to the special combination $\sqrt{s}\rho^1_{j,H}(s)+\rho^0_{j,H}(s)$. On the other hand, we
can obtain the QCD sum rules for the  charmed baryon states with negative parity,
 \begin{eqnarray}
\int_{m_c^2}^{s_0}ds \left[\sqrt{s}\rho^1_{j,H}(s)-\rho^0_{j,H}(s)\right]\exp\left( -\frac{s}{T^2}\right)
&=&2M_{-}{\lambda^{-}_{j}}^2\exp\left( -\frac{M_{-}^2}{T^2}\right) \, .
\end{eqnarray}
The contributions of the ${\frac{3}{2}}^{\pm}$ and ${\frac{5}{2}}^{\pm}$ charmed baryon states can be separated  unambiguously. In this article, we will focus on the $\Lambda_c$ and $\Xi_c$ states with positive parity.

At the QCD side, we  calculate the light quark parts of the correlation functions
 $\Pi_{\alpha\beta}(p)$ and $\Pi_{\alpha\beta\mu\nu}(p)$ with the full light quark propagators $S_{ij}(x)$  in the coordinate space
 \begin{eqnarray}
S_{ij}(x)&=& \frac{i\delta_{ij}\!\not\!{x}}{ 2\pi^2x^4}
-\frac{\delta_{ij}m_q}{4\pi^2x^2}-\frac{\delta_{ij}\langle
\bar{q}q\rangle}{12} +\frac{i\delta_{ij}\!\not\!{x}m_q
\langle\bar{q}q\rangle}{48}-\frac{\delta_{ij}x^2\langle \bar{q}g_s\sigma Gq\rangle}{192}\nonumber\\
&&+\frac{i\delta_{ij}x^2\!\not\!{x} m_q\langle \bar{q}g_s\sigma
 Gq\rangle }{1152} -\frac{ig_s G^{a}_{\alpha\beta}t^a_{ij}(\!\not\!{x}
\sigma^{\alpha\beta}+\sigma^{\alpha\beta} \!\not\!{x})}{32\pi^2x^2} -\frac{1}{8}\langle\bar{q}_j\sigma^{\mu\nu}q_i \rangle \sigma_{\mu\nu} +\cdots \, , \nonumber \\
\end{eqnarray}
 and take   the full $c$-quark propagator $C_{ij}(x)$ in the momentum space,
\begin{eqnarray}
C_{ij}(x)&=&\frac{i}{(2\pi)^4}\int d^4k e^{-ik \cdot x} \left\{
\frac{\delta_{ij}}{\!\not\!{k}-m_c}
-\frac{g_sG^n_{\alpha\beta}t^n_{ij}}{4}\frac{\sigma^{\alpha\beta}(\!\not\!{k}+m_c)+(\!\not\!{k}+m_c)
\sigma^{\alpha\beta}}{(k^2-m_c^2)^2}\right.\nonumber\\
&&\left. -\frac{g_s^2 (t^at^b)_{ij} G^a_{\alpha\beta}G^b_{\mu\nu}(f^{\alpha\beta\mu\nu}+f^{\alpha\mu\beta\nu}+f^{\alpha\mu\nu\beta}) }{4(k^2-m_c^2)^5}+\cdots\right\} \, , \end{eqnarray}
\begin{eqnarray}
f^{\alpha\beta\mu\nu}&=&(\!\not\!{k}+m_c)\gamma^\alpha(\!\not\!{k}+m_c)\gamma^\beta(\!\not\!{k}+m_c)\gamma^\mu(\!\not\!{k}+m_c)\gamma^\nu(\!\not\!{k}+m_c)\, ,
\end{eqnarray}
 $q=u,d,s$,  $t^n=\frac{\lambda^n}{2}$, the $\lambda^n$ is the Gell-Mann matrix \cite{PRT85}. In Eq.(27), we retain the term $\langle\bar{q}_j\sigma_{\mu\nu}q_i \rangle$  originates from the Fierz re-arrangement of the $\langle q_i \bar{q}_j\rangle$ to  absorb the gluons  emitted from the other  quark lines to form
$\langle\bar{q}_j g_s G^a_{\alpha\beta} t^a_{mn}\sigma_{\mu\nu} q_i \rangle$  to extract the mixed condensate  $\langle\bar{q}g_s\sigma G q\rangle$.
Then we compute  the integrals both in the coordinate space and momentum space  to obtain the correlation functions $\Pi_{j}(p^2)$, and  obtain  the QCD spectral densities  through  dispersion relation,
\begin{eqnarray}
\frac{{\rm Im}\Pi_{j}(s)}{\pi}&=&\!\not\!{p}\, \rho^1_{j,QCD}(s)+\rho^0_{j,QCD}(s) \, ,
\end{eqnarray}
where $j=\frac{3}{2}$, $\frac{5}{2}$, the explicit expressions of the QCD spectral densities $\rho^1_{j,QCD}(s)$ and $\rho^0_{j,QCD}(s)$ are shown in the Appendix. In this article, we carry out the operator product expansion up to the vacuum condensates of dimension 10
and take into account the condensates, which are    vacuum expectations
of the operators of order    $\mathcal{O}( \alpha_s^{k})$ with $k\leq 1$,  in a consistent way. In calculations, we observe that only
  the vacuum condensates $\langle\bar{q}q\rangle$, $\langle\bar{s}s\rangle$, $\langle \frac{\alpha_sGG}{\pi}\rangle$, $\langle\bar{q}g_s\sigma Gq\rangle$, $\langle\bar{s}g_s\sigma Gs\rangle$, $\langle\bar{q}g_s\sigma Gq\rangle^2$,  $ \langle\bar{q}g_s\sigma Gq\rangle\langle\bar{s}g_s\sigma Gs\rangle$ have contributions.

Once the analytical expressions of the QCD spectral densities $\rho^1_{j,QCD}(s)$ and $\rho^0_{j,QCD}(s)$ are obtained,  we  take the
quark-hadron duality below the continuum thresholds  $s_0$ and introduce the exponential function $\exp\left(-\frac{s}{T^2}\right)$ to depress the continuum state contributions to obtain  the  QCD sum rules:
\begin{eqnarray}
2M_{+}{\lambda^{+}_{j}}^2\exp\left( -\frac{M_{+}^2}{T^2}\right)
&=& \int_{m_c^2}^{s_0}ds \left[\sqrt{s}\rho^1_{j,QCD}(s)+\rho^0_{j,QCD}(s)\right]\exp\left( -\frac{s}{T^2}\right)\, .
\end{eqnarray}

We derive   Eq.(31) with respect to  $\frac{1}{T^2}$, then eliminate the
 pole residues $\lambda^{+}_j$ and obtain the QCD sum rules for
 the masses of the charmed baryon  states with $J^P={\frac{3}{2}}^+$ and ${\frac{5}{2}}^+$,
 \begin{eqnarray}
 M^2_{+} &=& \frac{-\frac{d}{d(1/T^2)}\int_{m_c^2}^{s_0}ds \left[\sqrt{s}\rho^1_{j,QCD}(s)+\rho^0_{j,QCD}(s)\right]\exp\left( -\frac{s}{T^2}\right)}{\int_{m_c^2}^{s_0}ds \left[\sqrt{s}\rho^1_{j,QCD}(s)+\rho^0_{j,QCD}(s)\right]\exp\left( -\frac{s}{T^2}\right)}\, .
\end{eqnarray}

\section{Numerical results and discussions}
The input parameters at the QCD side are taken to be the standard values
$\langle\bar{q}q \rangle=-(0.24\pm 0.01\, \rm{GeV})^3$,  $\langle\bar{s}s \rangle=(0.8\pm0.1)\langle\bar{q}q \rangle$,
 $\langle\bar{q}g_s\sigma G q \rangle=m_0^2\langle \bar{q}q \rangle$,  $\langle\bar{s}g_s\sigma G s \rangle=m_0^2\langle \bar{s}s \rangle$,
$m_0^2=(0.8 \pm 0.1)\,\rm{GeV}^2$, $\langle \frac{\alpha_s
GG}{\pi}\rangle=(0.33\,\rm{GeV})^4 $    at the energy scale  $\mu=1\, \rm{GeV}$
\cite{SVZ79,PRT85,ColangeloReview}, $m_{c}(m_c)=(1.275\pm0.025)\,\rm{GeV}$ and $m_s(\mu=2\,\rm{GeV})=(0.095\pm0.005)\,\rm{GeV}$
 from the Particle Data Group \cite{PDG}. Furthermore, we set $m_u=m_d=0$ due to the small current quark masses.
 We take into account
the energy-scale dependence of  the input parameters from the renormalization group equation,
\begin{eqnarray}
\langle\bar{q}q \rangle(\mu)&=&\langle\bar{q}q \rangle(Q)\left[\frac{\alpha_{s}(Q)}{\alpha_{s}(\mu)}\right]^{\frac{4}{9}}\, ,\nonumber\\
\langle\bar{s}s \rangle(\mu)&=&\langle\bar{s}s \rangle(Q)\left[\frac{\alpha_{s}(Q)}{\alpha_{s}(\mu)}\right]^{\frac{4}{9}}\, , \nonumber\\
\langle\bar{q}g_s \sigma Gq \rangle(\mu)&=&\langle\bar{q}g_s \sigma Gq \rangle(Q)\left[\frac{\alpha_{s}(Q)}{\alpha_{s}(\mu)}\right]^{\frac{2}{27}}\, , \nonumber\\
\langle\bar{s}g_s \sigma Gs \rangle(\mu)&=&\langle\bar{s}g_s \sigma Gs \rangle(Q)\left[\frac{\alpha_{s}(Q)}{\alpha_{s}(\mu)}\right]^{\frac{2}{27}}\, , \nonumber\\
m_c(\mu)&=&m_c(m_c)\left[\frac{\alpha_{s}(\mu)}{\alpha_{s}(m_c)}\right]^{\frac{12}{25}} \, ,\nonumber\\
m_s(\mu)&=&m_s({\rm 2GeV} )\left[\frac{\alpha_{s}(\mu)}{\alpha_{s}({\rm 2GeV})}\right]^{\frac{4}{9}} \, ,\nonumber\\
\alpha_s(\mu)&=&\frac{1}{b_0t}\left[1-\frac{b_1}{b_0^2}\frac{\log t}{t} +\frac{b_1^2(\log^2{t}-\log{t}-1)+b_0b_2}{b_0^4t^2}\right]\, ,
\end{eqnarray}
  where $t=\log \frac{\mu^2}{\Lambda^2}$, $b_0=\frac{33-2n_f}{12\pi}$, $b_1=\frac{153-19n_f}{24\pi^2}$, $b_2=\frac{2857-\frac{5033}{9}n_f+\frac{325}{27}n_f^2}{128\pi^3}$,  $\Lambda=213\,\rm{MeV}$, $296\,\rm{MeV}$  and  $339\,\rm{MeV}$ for the flavors  $n_f=5$, $4$ and $3$, respectively  \cite{PDG}, and evolve all the input parameters to the optimal energy scales  $\mu$ to extract the masses of the charmed baryon states.

  In the heavy quark limit, the $Q$-quark serves as a static well potential and  combines with a  light quark $q$  to form a heavy diquark  in  color antitriplet,
or combines with a  light antiquark $\bar{q}$ to form a heavy meson in color singlet (meson-like state in color octet), or combines with a light diquark $\varepsilon^{ijk}q^i\,q^{\prime j}$ to form a heavy baryon in color singlet (triquark in color triplet),
\begin{eqnarray}
q^j+Q^k &\to & \varepsilon^{ijk}\, q^j\,Q^k\, , \nonumber\\
\bar{q}^{ j}+Q^k &\to & \bar{q}^{ j} \,\delta_{jk}\, Q^k\,\, (\bar{q}^{ j}\,\lambda^{a}_{jk}\,Q^k) \, , \nonumber\\
\varepsilon^{ijl}q^i\,q^{\prime j} +Q^k&\to & \varepsilon^{ijl}\,q^i \,q^{\prime j}\, \delta_{lk}\,Q^k\,\,(\varepsilon^{lkm}\varepsilon^{ijl}\,q^i \,q^{\prime j}\, Q^k)\, ,
\end{eqnarray}
where the $i$, $j$, $k$, $l$, $m$ are color indexes, the $\lambda^a$ is Gell-Mann matrix.  The $\overline{Q}$-quark serves  as another static well potential and has similar  property.
 Then
\begin{eqnarray}
 \varepsilon^{ijk}\, q^j\,Q^k+\varepsilon^{imn} \bar{q}^{\prime m}\,\overline{Q}^n &\to &  {\rm compact \,\,\, tetraquark \,\,\, states}\, , \nonumber\\
  \varepsilon^{lkm}\varepsilon^{ijl}\,q^i \,q^{\prime j}\, Q^k+\varepsilon^{mnb} q^{\prime \prime n}\,\overline{Q}^b &\to &  {\rm compact \,\,\, pentaquark \,\,\, states}\, , \nonumber\\
 \varepsilon^{ijk}\,q^i \,q^{\prime j}\,Q^k+\overline{Q}q^{\prime\prime} &\to & {\rm loose  \,\,\, molecular \,\,\, states}\, , \nonumber\\
 \bar{q}Q+\overline{Q}q^{\prime} &\to & {\rm loose  \,\,\, molecular \,\,\, states}\, , \nonumber\\
  \bar{q}\lambda^aQ+\overline{Q}\lambda^a q^{\prime} &\to & {\rm   molecule-like  \,\,\, states}\, .
\end{eqnarray}

The three-quark systems $qq^\prime Q$, four-quark systems $q\bar{q}^\prime Q \overline{Q}$, five-quark systems $qq^\prime q^{\prime\prime} Q \overline{Q}$    are characterized by the effective heavy quark masses ${\mathbb{M}}_Q$ (or constituent quark masses) and the virtuality  $V=\sqrt{M^2_{B}-{\mathbb{M}}_Q^2}$, $\sqrt{M^2_{X/Y/Z}-(2{\mathbb{M}}_Q)^2}$, $\sqrt{M^2_{P}-(2{\mathbb{M}}_Q)^2}$ (or bound energy not as robust), where the $B$ denotes the conventional baryon states, the $X$, $Y$, $Z$ denote the hidden-charm (bottom) tetraquark quark states, molecular states or molecule-like states, the $P$ denotes  the (molecular) pentaquark  states. It is natural to take the energy  scales of the QCD spectral densities to be $\mu=V$.

The effective $Q$-quark masses ${\mathbb{M}}_Q$ have three universal values, which correspond to\\
$\bf (1)$ the diquark-quark type baryon states $\varepsilon^{ijk}\,q^i \,q^{\prime j}\, \,Q^k$,\\
 the diquark-antidiquark type tetraquark states $ \varepsilon^{ijk}\varepsilon^{imn}\, q^j\,Q^k \bar{q}^{\prime m}\,\overline{Q}^n$, \\
 the diquark-diquark-antiquark type pentaquark states $ \varepsilon^{lkm}\varepsilon^{ijl}\varepsilon^{mnb}\,q^i \,q^{\prime j}\, Q^k \,q^{\prime \prime n}\,\overline{Q}^b$,\\
$\bf (2)$ the meson-meson type molecular states $\bar{q}Q\,\overline{Q}q^{\prime}$, \\
$\bf (3)$ the molecule-like states $\bar{q}\lambda^aQ\,\overline{Q}\lambda^a q^{\prime}$,\\
shown in Eqs.(34-35), respectively, and embody  the net effects of the complex dynamics \cite{WangPc,WangTetraquark,WangMolecule,WangIJMPA}.

We fit  the   effective $Q$-quark masses ${\mathbb{M}}_{Q}$ to reproduce the experimental values $M_{Z_c(3900)}$ and $M_{Z_b(10610)}$ in the  scenario of  tetraquark  states \cite{WangTetraquark}, then  we take the ${\mathbb{M}}_{c}$ and ${\mathbb{M}}_{b}$ as input parameters, and use the  energy scale formula $\mu=\sqrt{M^2_{X/Y/Z}-(2{\mathbb{M}}_Q)^2}$, $\sqrt{M^2_{P}-(2{\mathbb{M}}_c)^2}$, $\sqrt{M^2_{B}-{\mathbb{M}}_c^2}$ to study the  hidden-charm (hidden-bottom) tetraquark states,  hidden-charm pentaquark states and charmed  baryon states. We call the energy scale formula empirical,  because  the energy scale formula was used to study the  hidden-charm (hidden-bottom) tetraquark states and molecular states firstly \cite{WangTetraquark,WangMolecule}, then it was extended to study the hidden-charm pentaquark states \cite{WangPc} and charmed  baryon states \cite{Wang-2625-2815,WangOmega}.   The energy scale formula  works  well  for  the  $X(3872)$, $Z_c(3885/3900)$, $X^*(3860)$, $Y(3915)$ $Z_c(4020/4025)$,   $Z(4430)$, $X(4500)$, $Y(4630/4660)$, $X(4700)$, $Z_b(10610)$, $Z_b(10650)$ \cite{Wang-Y4274}\footnote{All the relevant references can be found in Ref.\cite{Wang-Y4274}. }, $P_c(4380)$, $P_c(4450)$ \cite{WangPc}, $\Lambda_c(2625)$, $\Xi_c(2815)$ \cite{Wang-2625-2815}, $\Omega_c(3050)$, $\Omega_c(3066)$, $\Omega_c(3090)$ and $\Omega_c(3119)$ \cite{WangOmega}.

   In this article,  we use the empirical  formula $ \mu =\sqrt{M_{B}^2-{\mathbb{M}}_c^2}$ to determine the ideal energy scales of the QCD spectral densities. If we take the updated value of the effective $c$-quark mass ${\mathbb{M}}_c=1.82\,\rm{GeV}$ \cite{WangEPJC4260}, then the optimal energy scales are $\mu=2.2\,\rm{GeV}$, $2.2\,\rm{GeV}$, $2.5\,\rm{GeV}$ and $2.5\,\rm{GeV}$ for the $\Lambda_c(2860)$, $\Lambda_c(2880)$, $\Xi_c(3055)$  and $\Xi_c(3080)$, respectively. In calculations, we observe that if the charmed baryon states $\Lambda_c(2860)$, $\Lambda_c(2880)$, $\Xi_c(3055)$  and $\Xi_c(3080)$ have the quantum numbers $(L_\rho,L_\lambda)=(0,2)$, the experimental values of the masses $M_{\Lambda_c/\Xi_c}$ can be reproduced approximately. The currents with the quantum numbers $(L_\rho,L_\lambda)=(2,0)$ and $(L_\rho,L_\lambda)=(1,1)$  couple potentially to the D-wave charmed baryon states  having  larger masses than the corresponding charmed baryon states $\Lambda_c(2860/2880)$ and $\Xi_c(3055/3080)$, so their QCD spectral densities  should be calculated at larger energy scales according to the virtuality  $V=\sqrt{M^2_{B}-{\mathbb{M}}_c^2}$, the empirical energy scale formula $ \mu =\sqrt{M_{B}^2-{\mathbb{M}}_c^2}$ serves  as a powerful constraint to satisfy. In Fig.1, we plot the masses  of the charmed baryon states $\Xi_c\left(0,2;\frac{5}{2}\right)$, $\Xi_c\left(0,2;\frac{3}{2}\right)$, $\Lambda_c\left(0,2;\frac{5}{2}\right)$ and $\Lambda_c\left(0,2;\frac{3}{2}\right)$ with variations of the energy scale $\mu$ for the central values of the Borel parameters  and threshold parameters shown in Table 1. From the figure, we can see that the predicted masses depend on the energy scale $\mu$ slightly, the acceptable ranges of the energy scale are rather large, the constraint $ \mu =\sqrt{M_{B}^2-{\mathbb{M}}_c^2}$ is not difficult  to satisfy in the present case. On the other hand, the pole residues increase monotonously and quickly with increase  of the energy scale, it is important to choose the ideal energy scales.

   We search for the ideal Borel parameters $T^2$ and continuum threshold parameters $s_0$   according to  the  four criteria:

$\bf{1_\cdot}$ Pole dominance at the hadron side, the pole contributions are about $(50-80)\%$;

$\bf{2_\cdot}$ Convergence of the operator product expansion, the dominant contributions come from the perturbative terms;

$\bf{3_\cdot}$ Appearance of the Borel platforms, the uncertainties $\delta M/M$ originate from the Borel parameters are about $(2-5)\%$ in the Borel windows;

$\bf{4_\cdot}$ Satisfying the energy scale formula. \\
by try and error,  and present the optimal energy scales $\mu$,  ideal Borel parameters $T^2$, continuum threshold parameters $s_0$,  pole contributions and perturbative contributions in Table 1.  In the QCD sum rules for the baryon states, the predicted masses usually increase  monotonously but slowly  with increase of the Borel parameters \cite{LiuYL}, there cannot appear platforms as flat as that appear in the case of the conventional mesons and tetraquark states \cite{ColangeloReview,WangTetraquark}. In this article, we observe that the predicted masses also increase  with increase of the Borel parameters,
so we constrain the  uncertainties $\delta M/M$ originate from the Borel parameters will not exceed  $5\%$ in the Borel windows.

From Table 1, we can see that the pole dominance at the hadron side is well satisfied and the operator product expansion is well convergent, the criteria $\bf{1}$ and $\bf{2}$ (the basic criteria of the QCD sum rules) are satisfied, so we expect to make reliable predictions. In Ref.\cite{Zhu-D-wave},  Chen et al study the  D-wave heavy baryon states with the QCD sum rules combined with the heavy quark effective theory, and extract the masses with the pole contributions  $\leq 20\%$, while in the present work, the pole contributions are about $(50-80)\%$.
   The QCD spectral densities have the terms $m_s\langle\bar{q}q\rangle$, $m_s\langle\bar{s}s\rangle$,  $m_s\langle\bar{q}g_s\sigma Gq\rangle$, $m_s\langle\bar{s}g_s\sigma Gs\rangle$, which are greatly depressed by the small $s$-quark mass and are of minor importance, the dominant contributions come from the perturbative terms.

\begin{table}
\begin{center}
\begin{tabular}{|c|c|c|c|c|c|c|c|}\hline\hline
             &$(L_\rho,L_\lambda)$  &$J^P$             &$\mu(\rm GeV)$ &$T^2 (\rm{GeV}^2)$  & $\sqrt{s_0} (\rm{GeV})$ & pole         & perturbative   \\  \hline
$\Xi_c$      &(0,2)                 &${\frac{5}{2}}^+$ &2.5            &$1.8-2.2$           & $3.7\pm0.1$             & $(46-76)\%$  & $(87-92)\%$  \\ \hline
$\Xi_c$      &(0,2)                 &${\frac{3}{2}}^+$ &2.5            &$1.5-1.9$           & $3.6\pm0.1$             & $(48-81)\%$  & $(96-99)\%$   \\ \hline
$\Lambda_c$  &(0,2)                 &${\frac{5}{2}}^+$ &2.2            &$1.5-1.9$           & $3.6\pm0.1$             & $(53-86)\%$  & $(76-90)\%$  \\ \hline
$\Lambda_c$  &(0,2)                 &${\frac{3}{2}}^+$ &2.2            &$1.2-1.6$           & $3.4\pm0.1$             & $(49-87)\%$  & $(88-99)\%$ \\ \hline

$\Xi_c$      &(2,0)                 &${\frac{5}{2}}^+$ &2.7            &$1.8-2.2$           & $3.8\pm0.1$             & $(47-77)\%$  & $(97-98)\%$   \\ \hline
$\Xi_c$      &(2,0)                 &${\frac{3}{2}}^+$ &2.7            &$1.7-2.1$           & $3.8\pm0.1$             & $(51-81)\%$  & $(98-99)\%$  \\ \hline
$\Lambda_c$  &(2,0)                 &${\frac{5}{2}}^+$ &2.7            &$1.7-2.1$           & $3.8\pm0.1$             & $(52-81)\%$  & $(95-97)\%$  \\ \hline
$\Lambda_c$  &(2,0)                 &${\frac{3}{2}}^+$ &2.7            &$1.7-2.1$           & $3.8\pm0.1$             & $(51-81)\%$  & $(96-97)\%$  \\ \hline

$\Xi_c$      &(1,1)                 &${\frac{5}{2}}^+$ &2.7            &$1.8-2.2$           & $3.8\pm0.1$             & $(50-79)\%$  & $(97-98)\%$ \\ \hline
$\Xi_c$      &(1,1)                 &${\frac{3}{2}}^+$ &2.7            &$1.6-2.0$           & $3.8\pm0.1$             & $(55-84)\%$  & $(101-102)\%$  \\ \hline
$\Lambda_c$  &(1,1)                 &${\frac{5}{2}}^+$ &2.7            &$1.8-2.2$           & $3.8\pm0.1$             & $(50-79)\%$  & $(96-97)\%$  \\ \hline
$\Lambda_c$  &(1,1)                 &${\frac{3}{2}}^+$ &2.7            &$1.6-2.0$           & $3.8\pm0.1$             & $(55-84)\%$  & $(100-100)\%$  \\ \hline\hline
\end{tabular}
\end{center}
\caption{ The optimal energy scales $\mu$, Borel parameters $T^2$, continuum threshold parameters $s_0$,
 pole contributions (pole)   and  perturbative contributions (perturbative) for the D-wave charmed baryon states.}
\end{table}

\begin{table}
\begin{center}
\begin{tabular}{|c|c|c|c|c|c|c|c|}\hline\hline
             &$(L_\rho,L_\lambda)$ &$J^P$              &$M(\rm{GeV})$           &$\lambda (\rm{GeV}^5)$               &(expt) (MeV)    &\cite{Zhu-D-wave} (GeV) \\ \hline
$\Xi_c$      &(0,2)                &${\frac{5}{2}}^+$  &$3.09^{+0.13}_{-0.15}$  &$3.73^{+0.89}_{-0.85}\times 10^{-2}$ &3076.94/3079.9  &$3.05^{+0.15}_{-0.16}$ \\ \hline
$\Xi_c$      &(0,2)                &${\frac{3}{2}}^+$  &$3.06^{+0.11}_{-0.13}$  &$1.47^{+0.37}_{-0.35}\times 10^{-1}$ &3055.1          &$3.04^{+0.15}_{-0.15}$\\ \hline
$\Lambda_c$  &(0,2)                &${\frac{5}{2}}^+$  &$2.88^{+0.18}_{-0.29}$  &$2.47^{+0.89}_{-0.92}\times 10^{-2}$ &2881.5          &$2.84^{+0.37}_{-0.20}$\\ \hline
$\Lambda_c$  &(0,2)                &${\frac{3}{2}}^+$  &$2.83^{+0.15}_{-0.24}$  &$0.84^{+0.32}_{-0.33}\times 10^{-1}$ &2856.1          &$2.81^{+0.33}_{-0.18}$ \\ \hline

$\Xi_c$      &(2,0)                &${\frac{5}{2}}^+$  &$3.25^{+0.10}_{-0.11}$  &$1.42^{+0.31}_{-0.27}\times 10^{-1}$ &                &$3.26^{+0.17}_{-0.15}$ \\ \hline
$\Xi_c$      &(2,0)                &${\frac{3}{2}}^+$  &$3.23^{+0.10}_{-0.11}$  &$2.50^{+0.56}_{-0.50}\times 10^{-1}$ &                &$3.25^{+0.16}_{-0.14}$\\ \hline
$\Lambda_c$  &(2,0)                &${\frac{5}{2}}^+$  &$3.22^{+0.10}_{-0.12}$  &$1.37^{+0.30}_{-0.28}\times 10^{-1}$ &                &$3.28^{+1.83}_{-0.30}$\\ \hline
$\Lambda_c$  &(2,0)                &${\frac{3}{2}}^+$  &$3.22^{+0.11}_{-0.11}$  &$2.50^{+0.56}_{-0.51}\times 10^{-1}$ &                &$3.25^{+1.72}_{-0.28}$\\ \hline

$\Xi_c$      &(1,1)                &${\frac{5}{2}}^+$  &$3.23^{+0.11}_{-0.11}$  &$6.02^{+1.22}_{-1.09}\times 10^{-2}$ &                &\\ \hline
$\Xi_c$      &(1,1)                &${\frac{3}{2}}^+$  &$3.22^{+0.10}_{-0.11}$  &$1.53^{+0.35}_{-0.31}\times 10^{-1}$ &                &\\ \hline
$\Lambda_c$  &(1,1)                &${\frac{5}{2}}^+$  &$3.23^{+0.10}_{-0.11}$  &$6.01^{+1.22}_{-1.09}\times 10^{-2}$ &                &\\ \hline
$\Lambda_c$  &(1,1)                &${\frac{3}{2}}^+$  &$3.21^{+0.11}_{-0.11}$  &$1.53^{+0.35}_{-0.32}\times 10^{-1}$ &                &\\ \hline\hline
\end{tabular}
\end{center}
\caption{ The masses and pole residues of the D-wave charmed baryon states, the masses are compared with the experimental data and other QCD sum rules predictions.
 }
\end{table}

\begin{figure}
 \centering
 \includegraphics[totalheight=6cm,width=8cm]{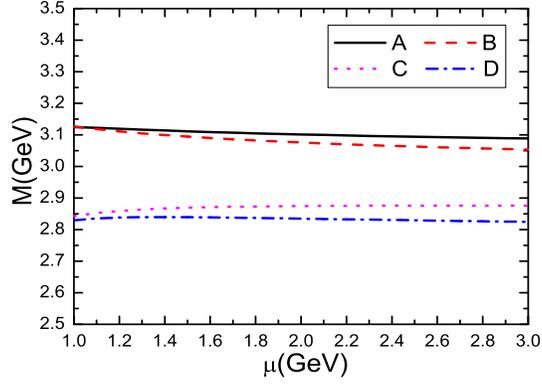}
         \caption{ The masses  of the charmed baryon states  with variations of the energy scale $\mu$ for the central values of the Borel parameters  and threshold parameters shown in Table 1, where the $A$, $B$, $C$ and $D$ correspond to the charmed baryon states $\Xi_c\left(0,2;\frac{5}{2}\right)$, $\Xi_c\left(0,2;\frac{3}{2}\right)$, $\Lambda_c\left(0,2;\frac{5}{2}\right)$ and $\Lambda_c\left(0,2;\frac{3}{2}\right)$, respectively.  }
\end{figure}

\begin{figure}
 \centering
 \includegraphics[totalheight=5cm,width=7cm]{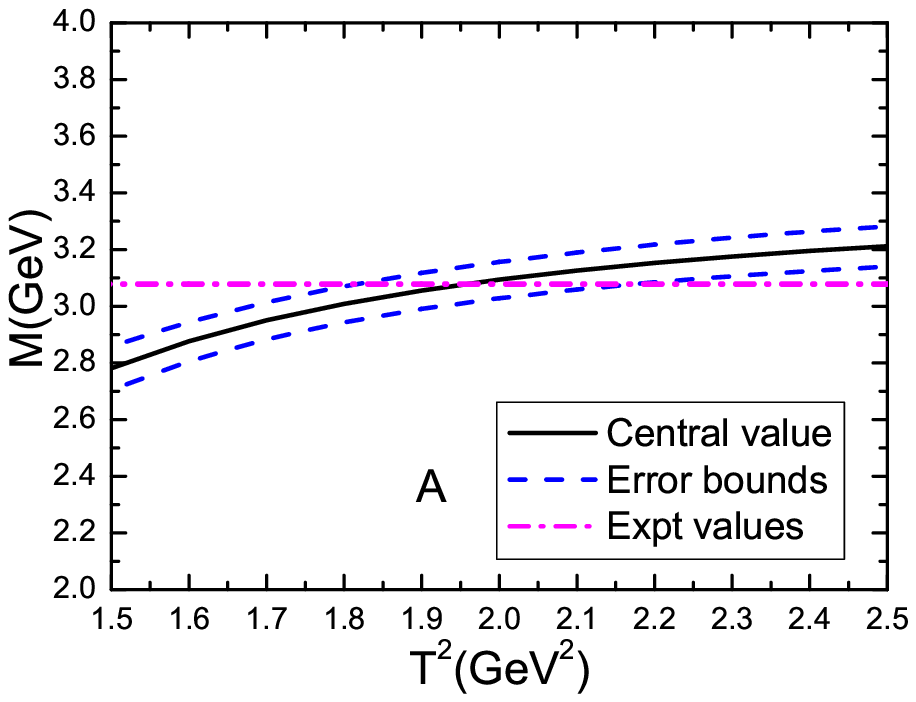}
 \includegraphics[totalheight=5cm,width=7cm]{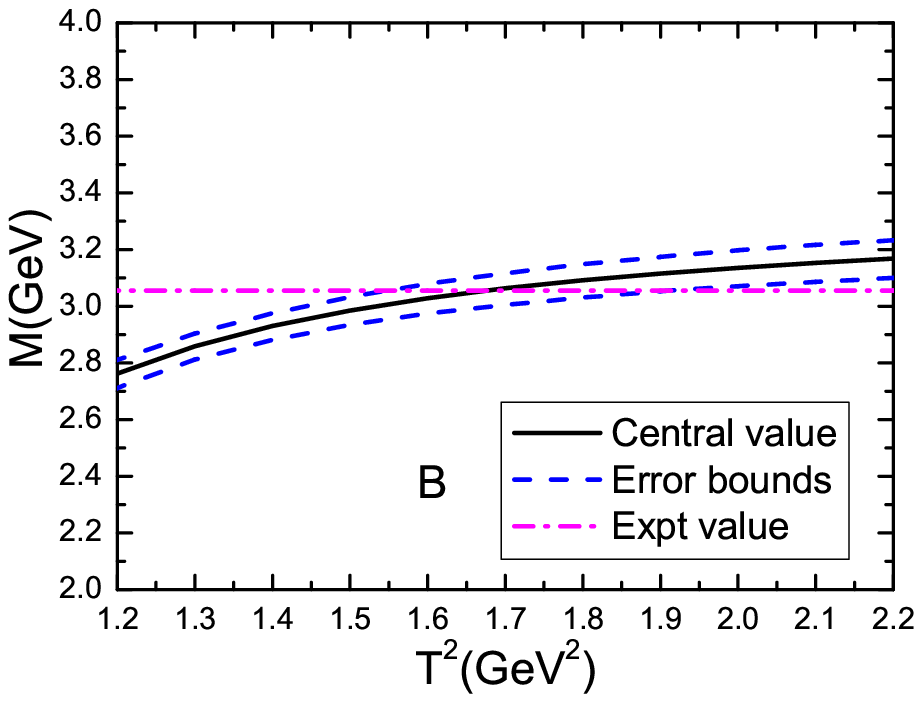}
 \includegraphics[totalheight=5cm,width=7cm]{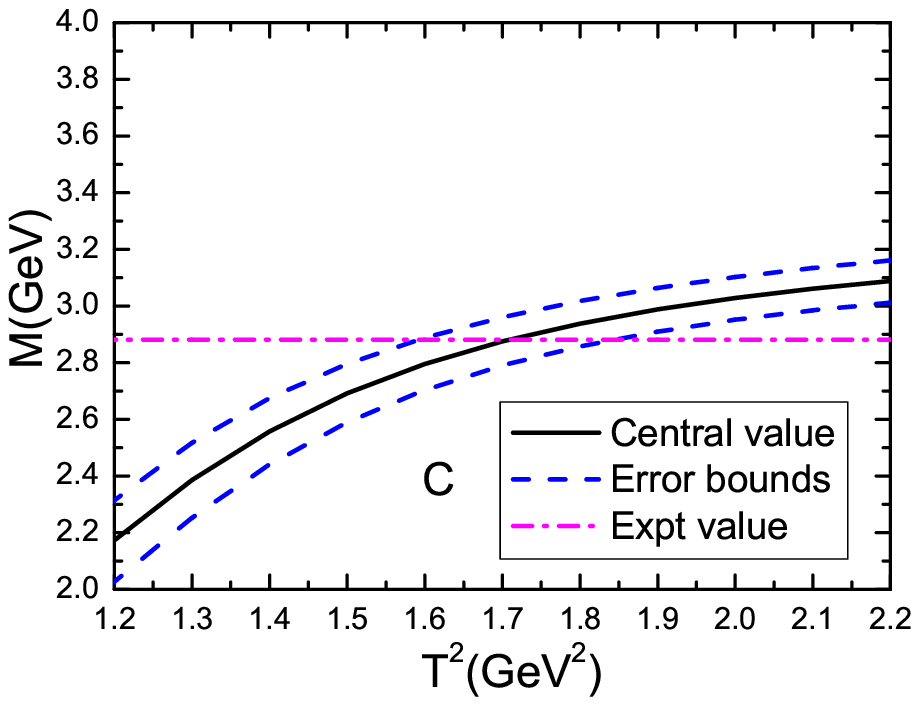}
 \includegraphics[totalheight=5cm,width=7cm]{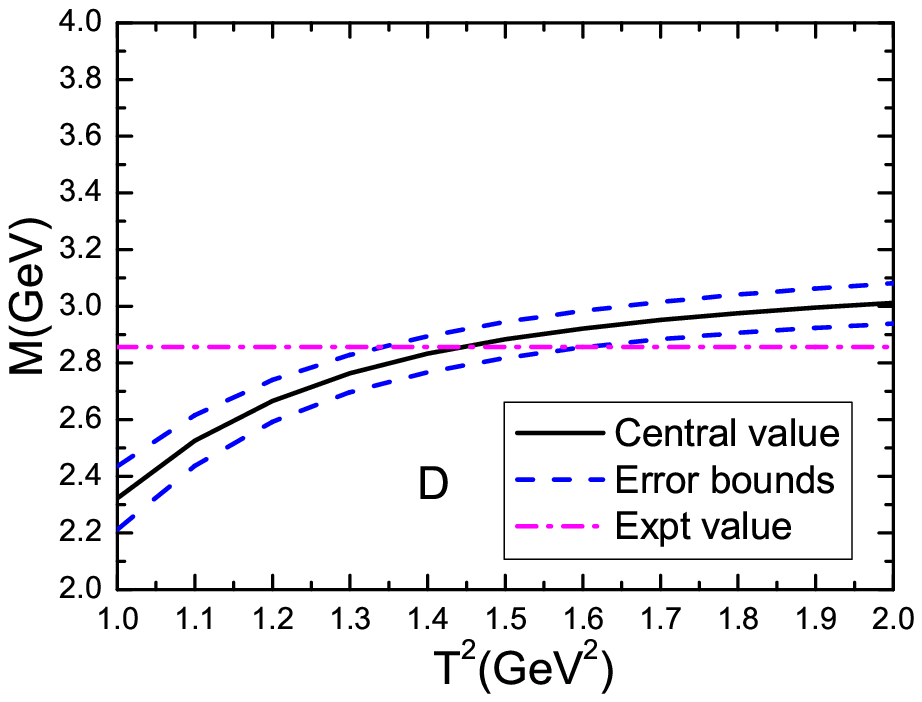}
        \caption{ The masses  of the charmed baryon states  with variations of the Borel parameters $T^2$, where the $A$, $B$, $C$ and $D$ correspond to the charmed baryon states $\Xi_c\left(0,2;\frac{5}{2}\right)$, $\Xi_c\left(0,2;\frac{3}{2}\right)$, $\Lambda_c\left(0,2;\frac{5}{2}\right)$ and $\Lambda_c\left(0,2;\frac{3}{2}\right)$, respectively, the Expt value denotes the experimental values.  }
\end{figure}

\begin{figure}
 \centering
 \includegraphics[totalheight=5cm,width=7cm]{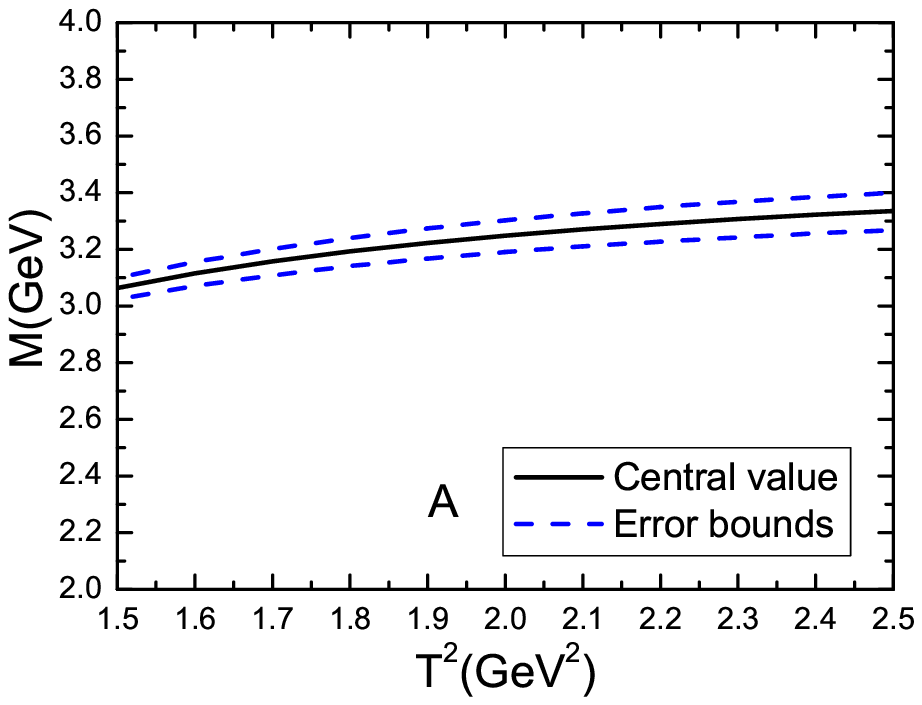}
 \includegraphics[totalheight=5cm,width=7cm]{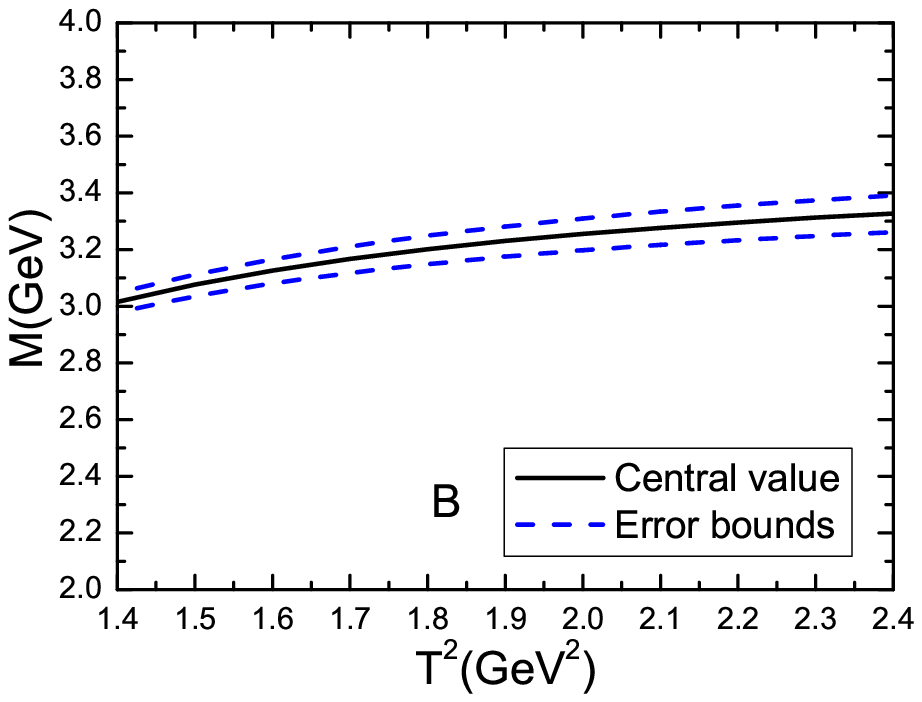}
 \includegraphics[totalheight=5cm,width=7cm]{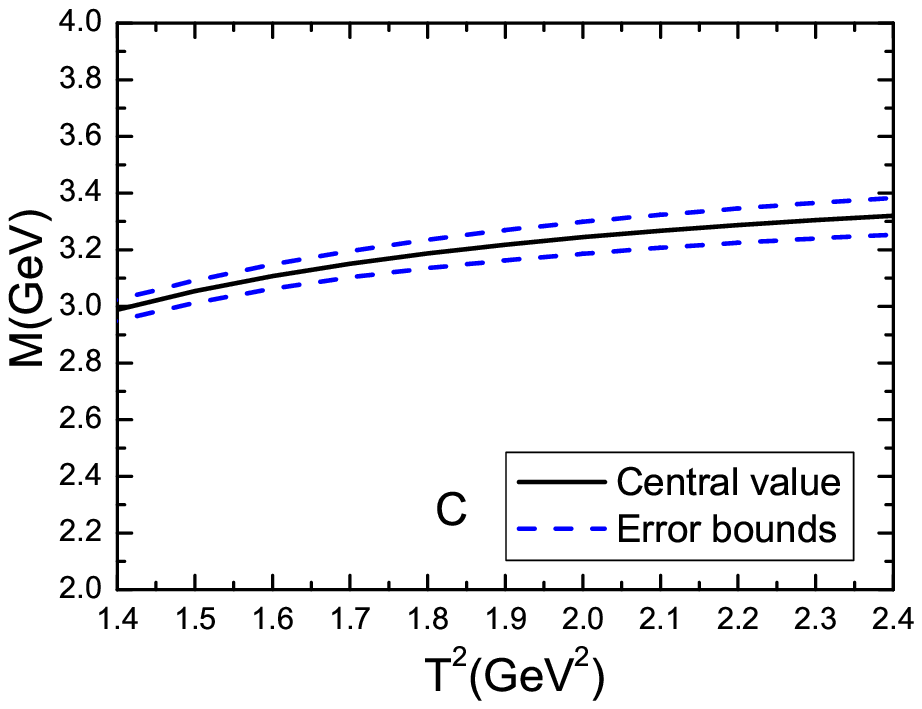}
 \includegraphics[totalheight=5cm,width=7cm]{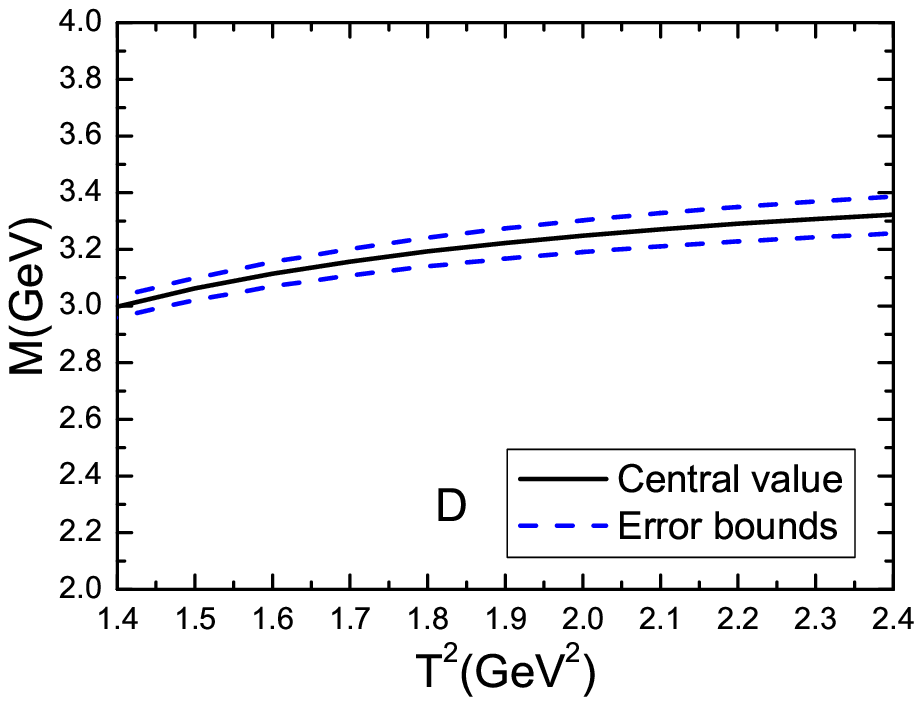}
        \caption{ The masses  of the charmed baryon states  with variations of the Borel parameters $T^2$, where the $A$, $B$, $C$ and $D$ correspond to the charmed baryon states $\Xi_c\left(2,0;\frac{5}{2}\right)$, $\Xi_c\left(2,0;\frac{3}{2}\right)$, $\Lambda_c\left(2,0;\frac{5}{2}\right)$ and $\Lambda_c\left(2,0;\frac{3}{2}\right)$, respectively.  }
\end{figure}

\begin{figure}
 \centering
 \includegraphics[totalheight=5cm,width=7cm]{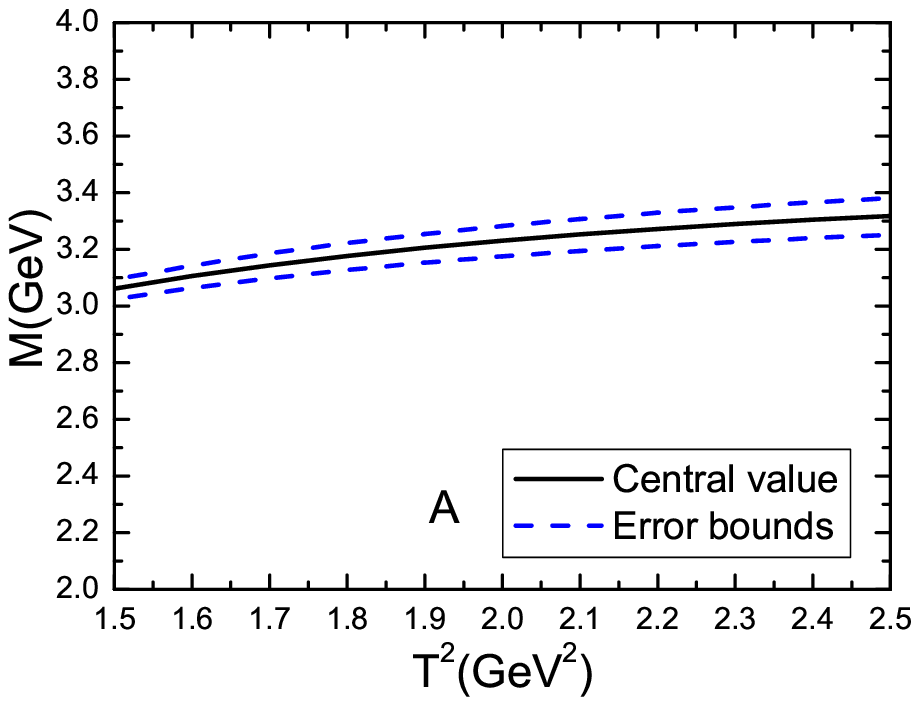}
 \includegraphics[totalheight=5cm,width=7cm]{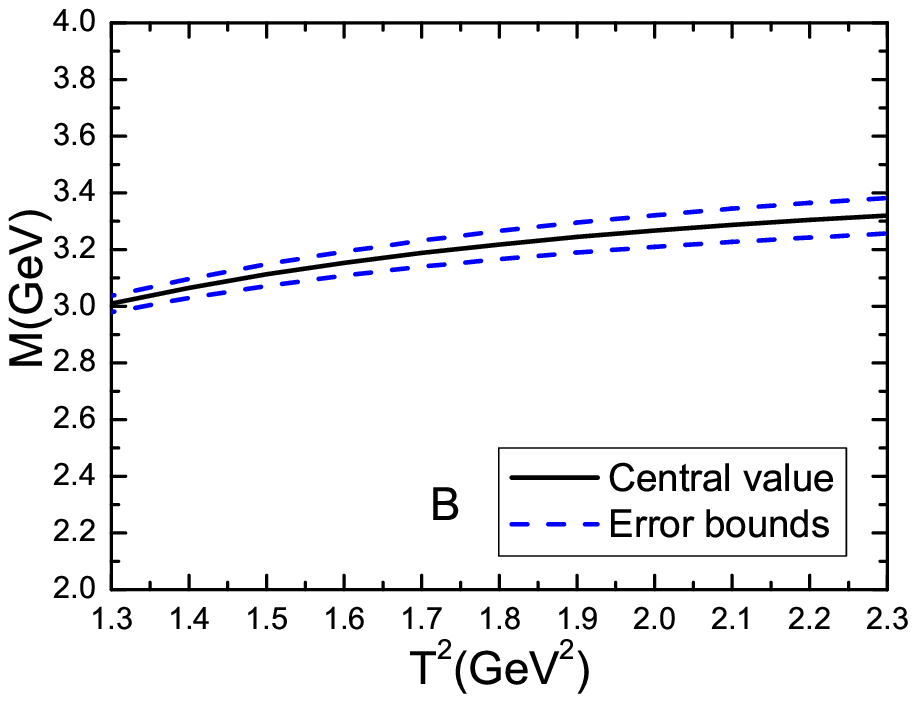}
 \includegraphics[totalheight=5cm,width=7cm]{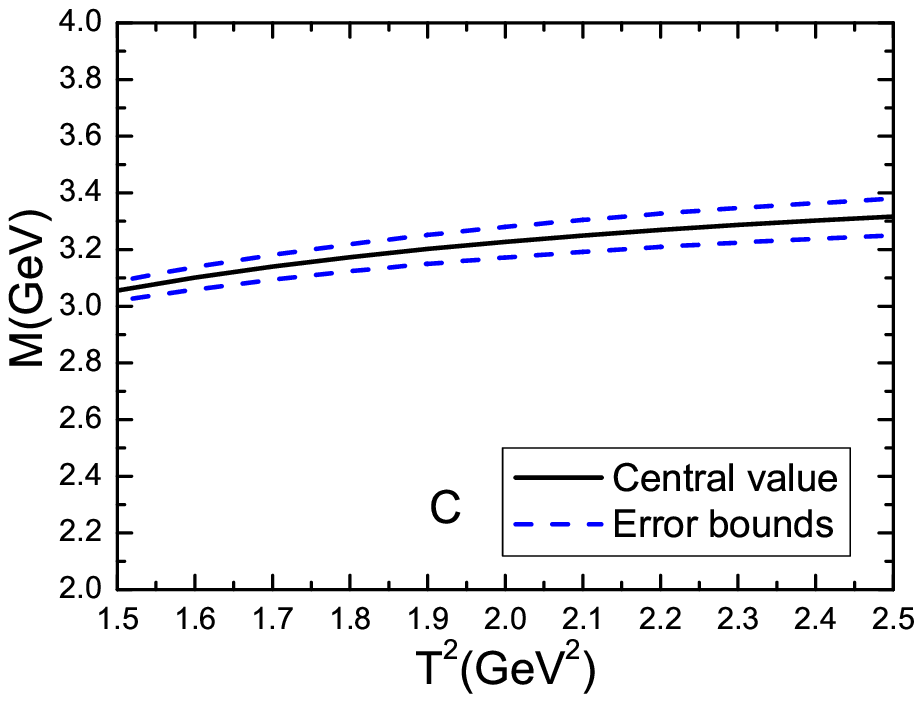}
 \includegraphics[totalheight=5cm,width=7cm]{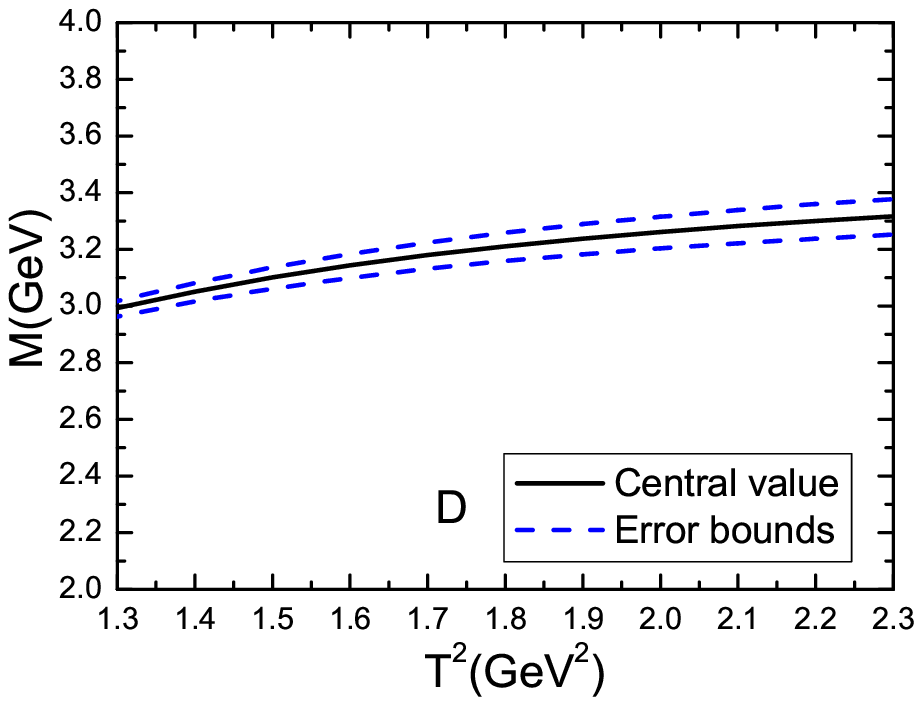}
        \caption{ The masses  of the charmed baryon states  with variations of the Borel parameters $T^2$, where the $A$, $B$, $C$ and $D$ correspond to the charmed baryon states $\Xi_c\left(1,1;\frac{5}{2}\right)$, $\Xi_c\left(1,1;\frac{3}{2}\right)$, $\Lambda_c\left(1,1;\frac{5}{2}\right)$ and $\Lambda_c\left(1,1;\frac{3}{2}\right)$, respectively.  }
\end{figure}

\begin{figure}
 \centering
 \includegraphics[totalheight=5cm,width=7cm]{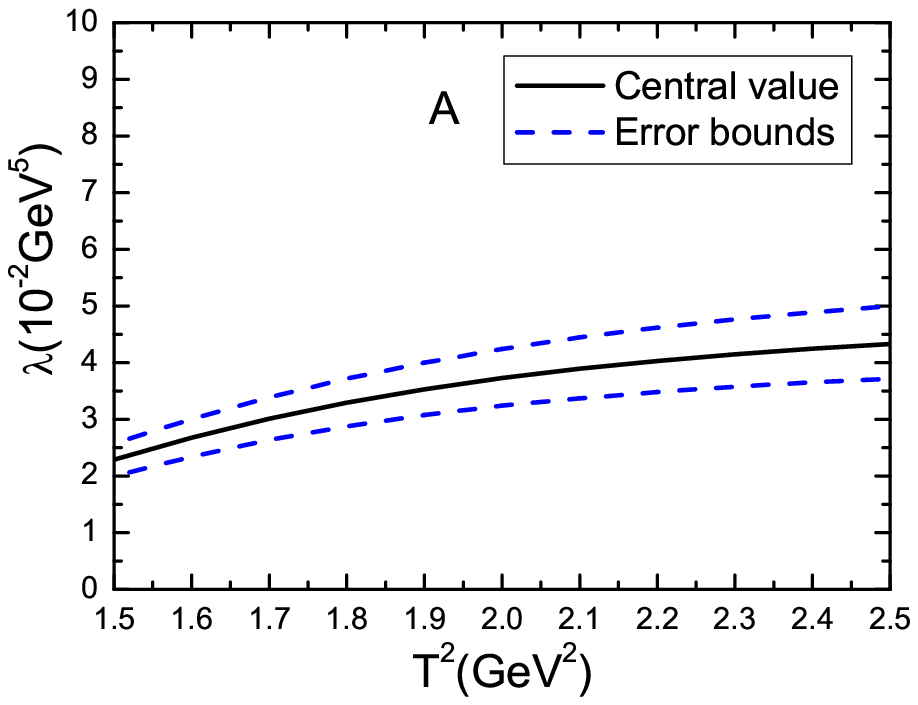}
 \includegraphics[totalheight=5cm,width=7cm]{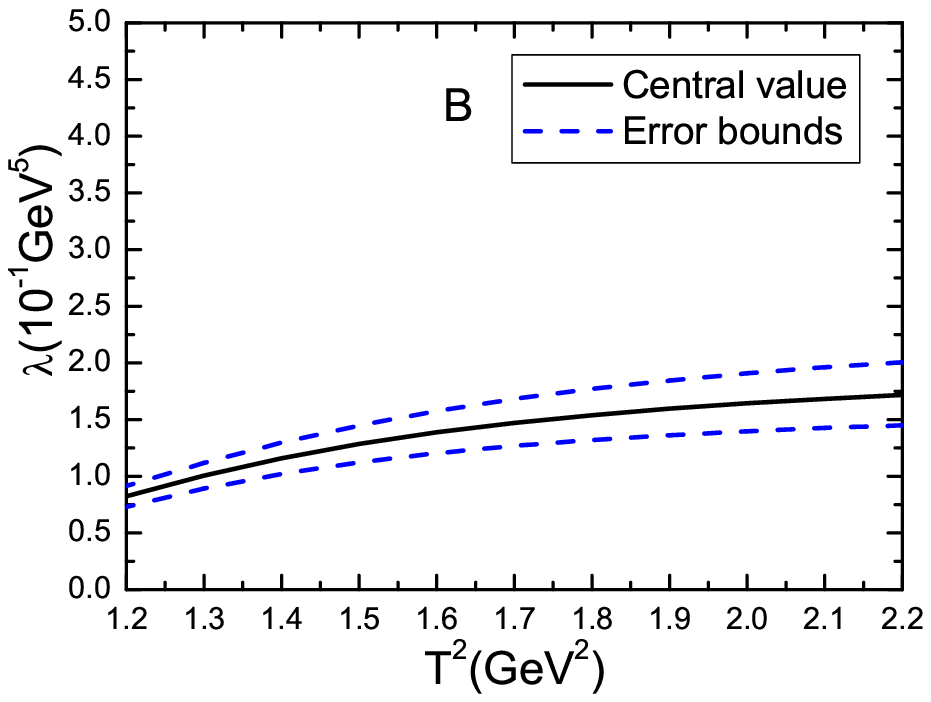}
 \includegraphics[totalheight=5cm,width=7cm]{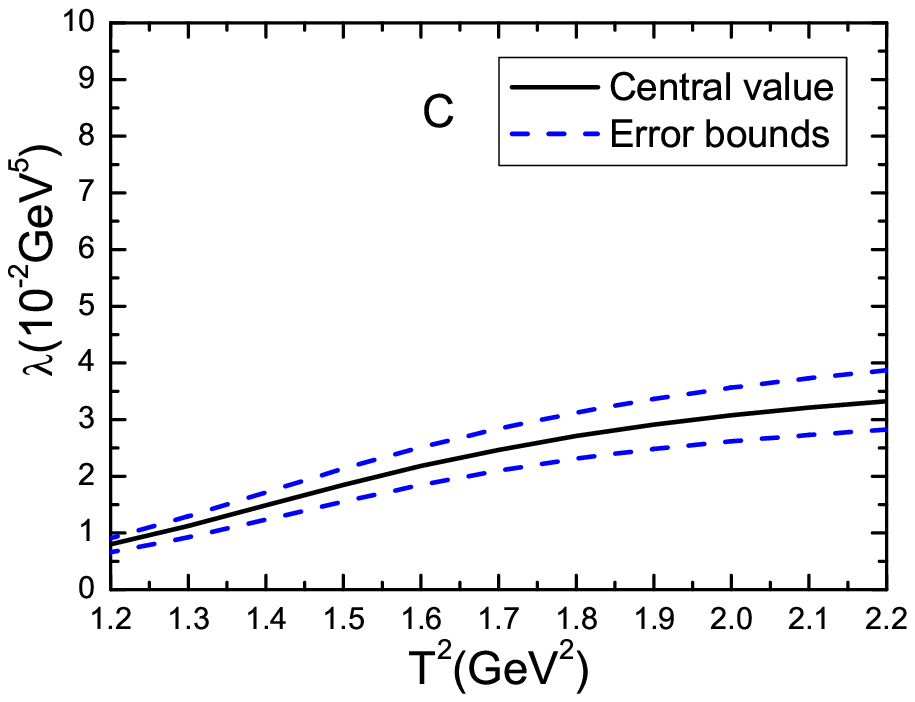}
 \includegraphics[totalheight=5cm,width=7cm]{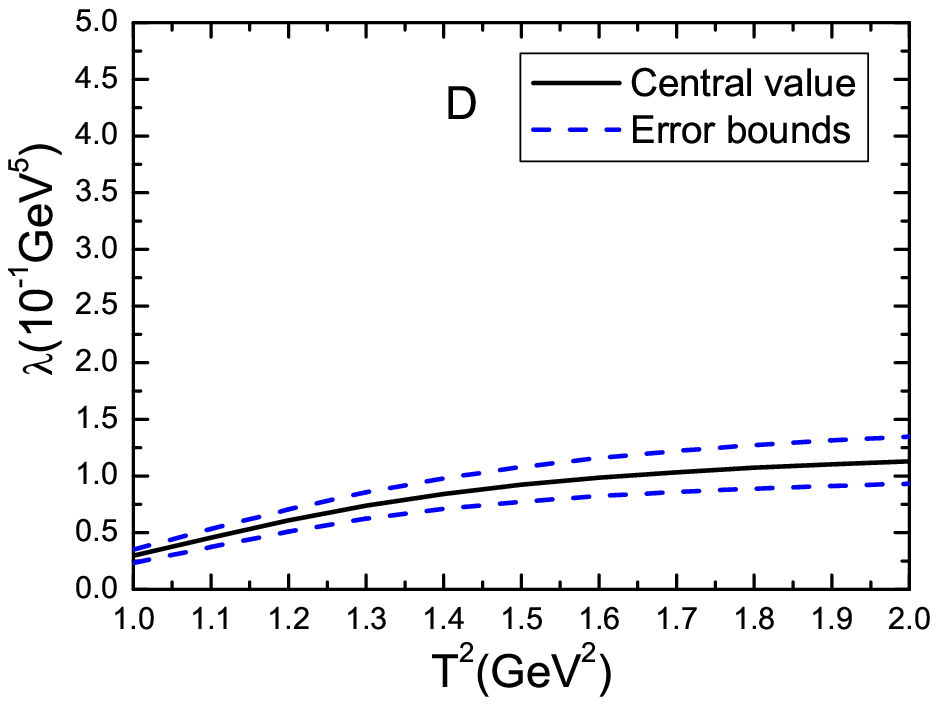}
        \caption{ The pole residues  of the charmed baryon states  with variations of the Borel parameters $T^2$, where the $A$, $B$, $C$ and $D$ correspond to the charmed baryon states $\Xi_c\left(0,2;\frac{5}{2}\right)$, $\Xi_c\left(0,2;\frac{3}{2}\right)$, $\Lambda_c\left(0,2;\frac{5}{2}\right)$ and $\Lambda_c\left(0,2;\frac{3}{2}\right)$, respectively. }
\end{figure}

\begin{figure}
 \centering
 \includegraphics[totalheight=5cm,width=7cm]{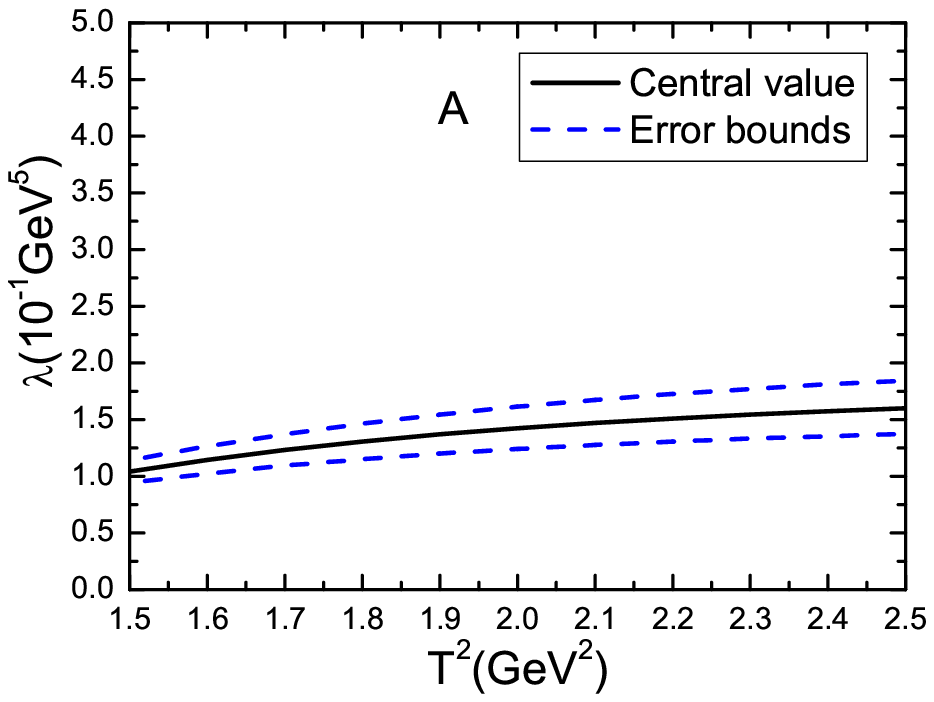}
 \includegraphics[totalheight=5cm,width=7cm]{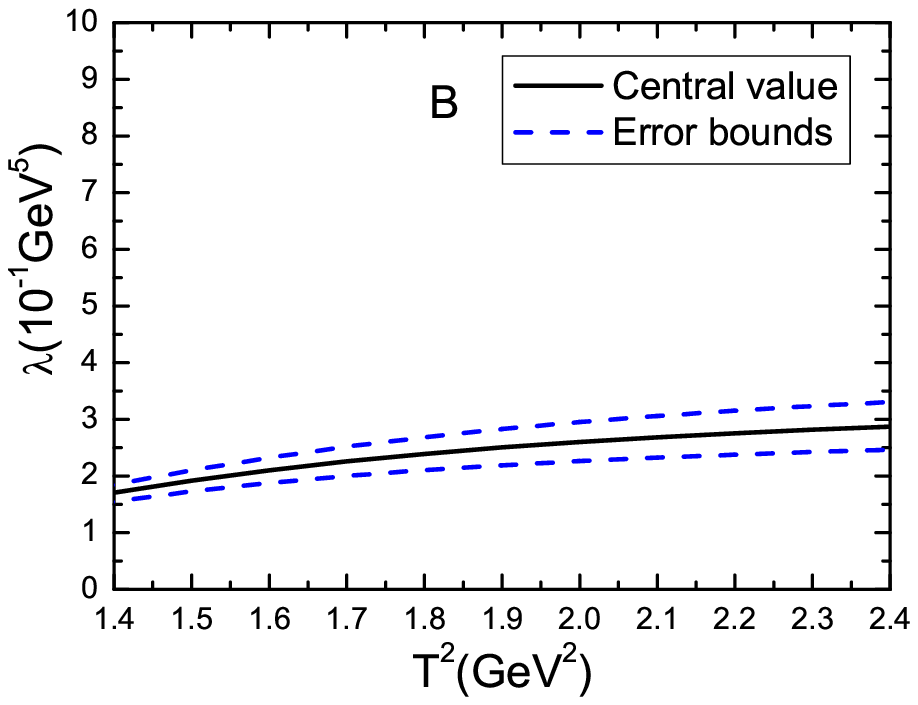}
 \includegraphics[totalheight=5cm,width=7cm]{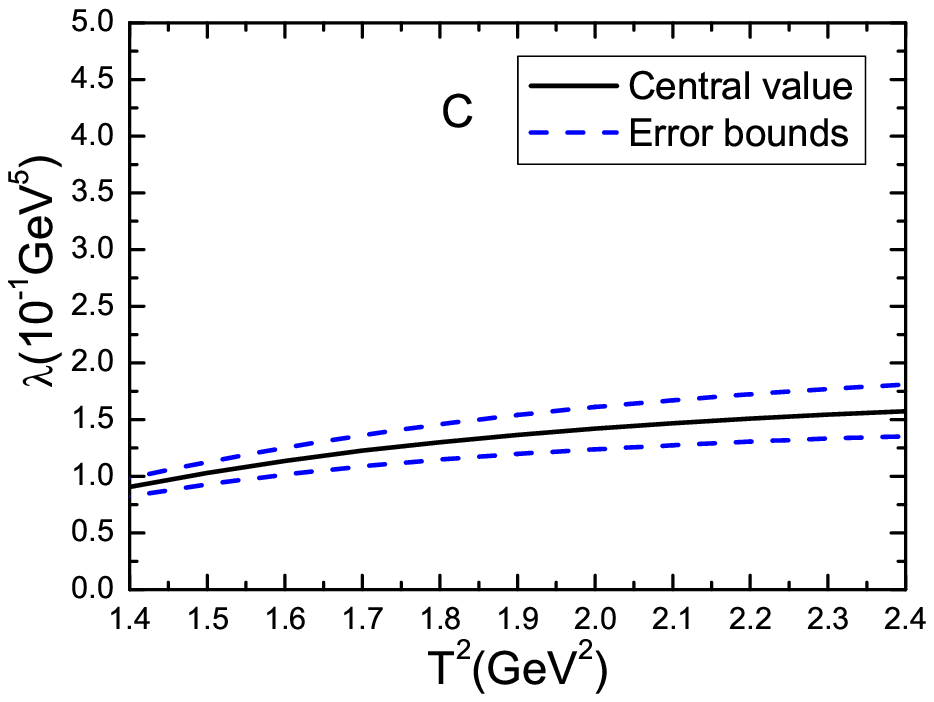}
 \includegraphics[totalheight=5cm,width=7cm]{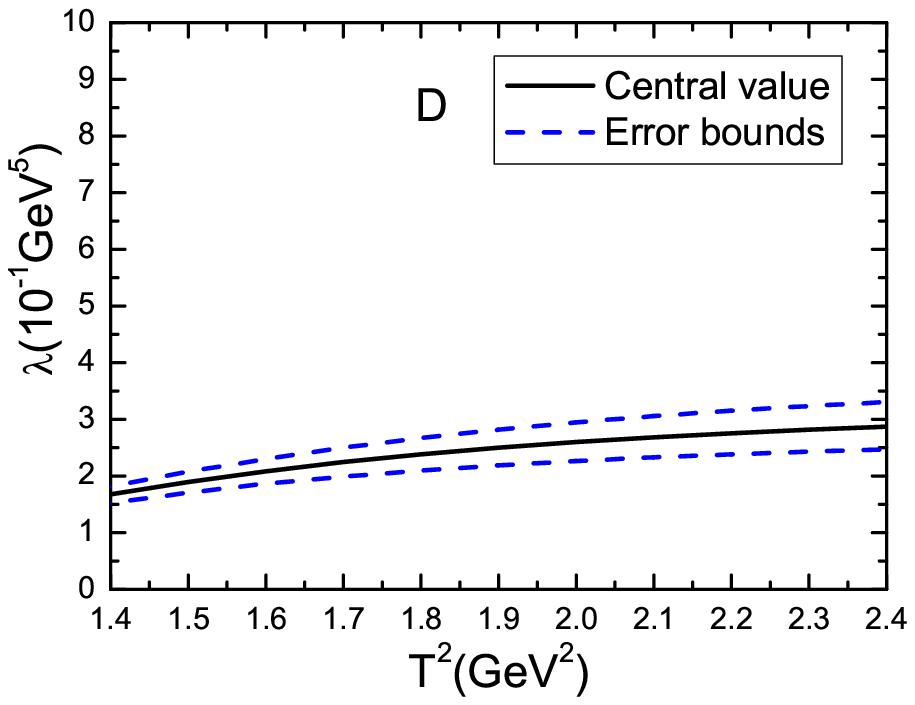}
        \caption{ The pole residues  of the charmed baryon states  with variations of the Borel parameters $T^2$, where the $A$, $B$, $C$ and $D$ correspond to the charmed baryon states $\Xi_c\left(2,0;\frac{5}{2}\right)$, $\Xi_c\left(2,0;\frac{3}{2}\right)$, $\Lambda_c\left(2,0;\frac{5}{2}\right)$ and $\Lambda_c\left(2,0;\frac{3}{2}\right)$, respectively.  }
\end{figure}

\begin{figure}
 \centering
 \includegraphics[totalheight=5cm,width=7cm]{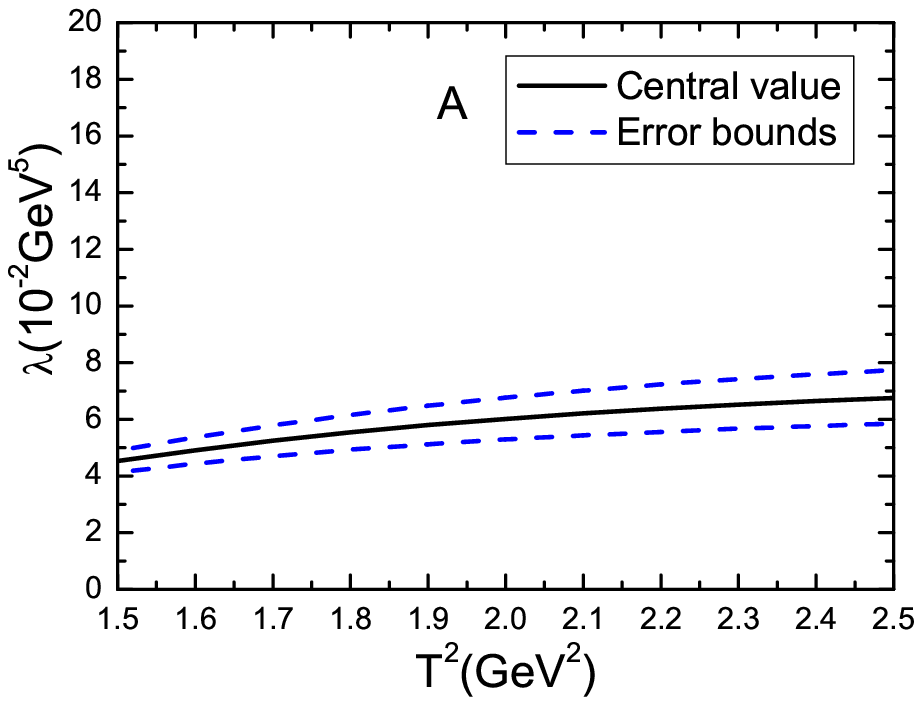}
 \includegraphics[totalheight=5cm,width=7cm]{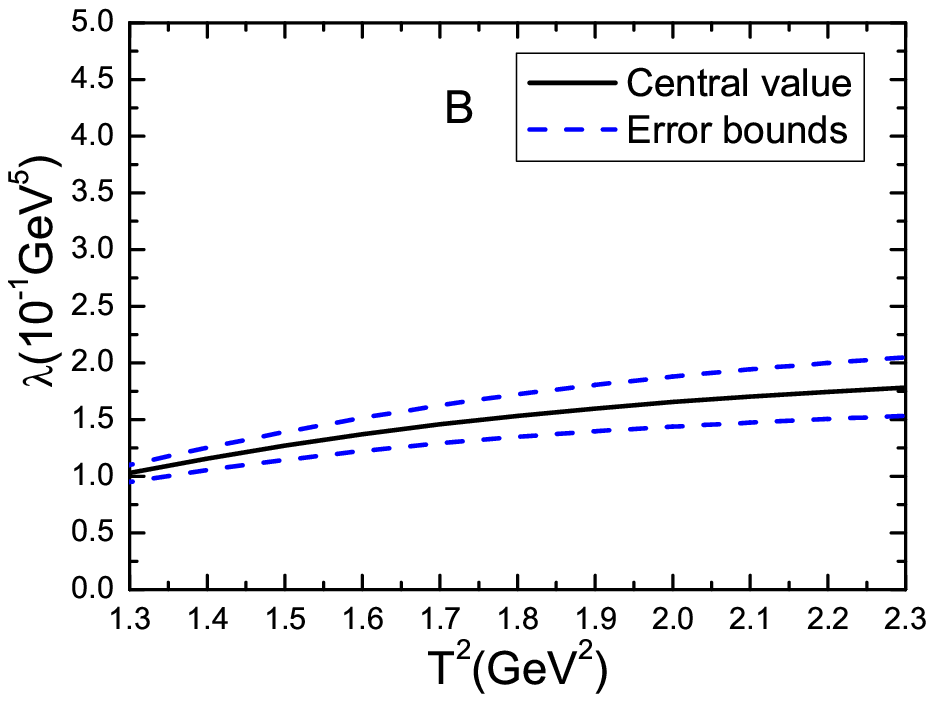}
 \includegraphics[totalheight=5cm,width=7cm]{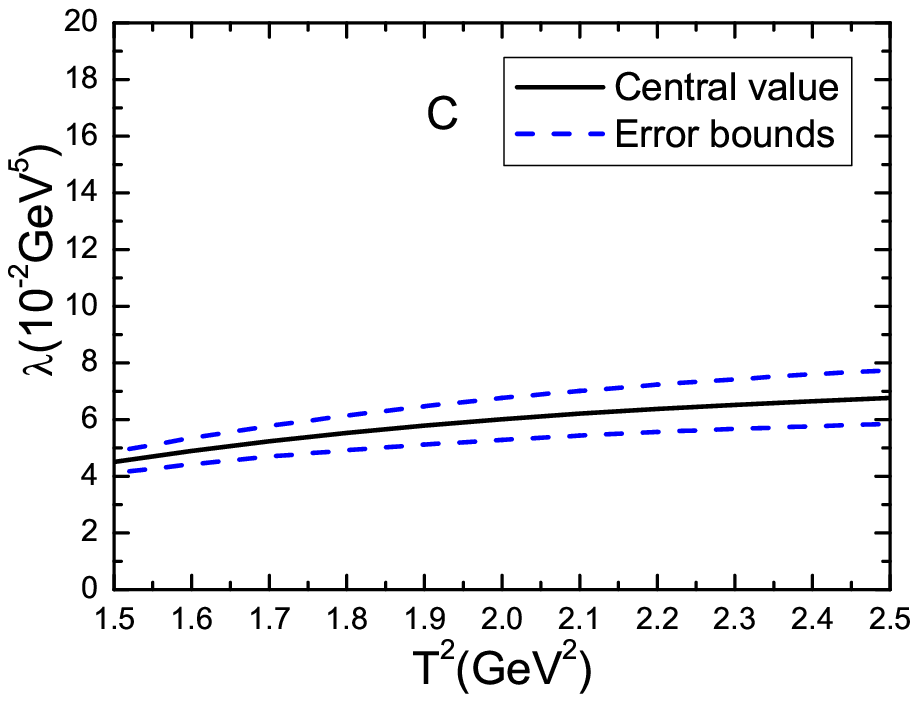}
 \includegraphics[totalheight=5cm,width=7cm]{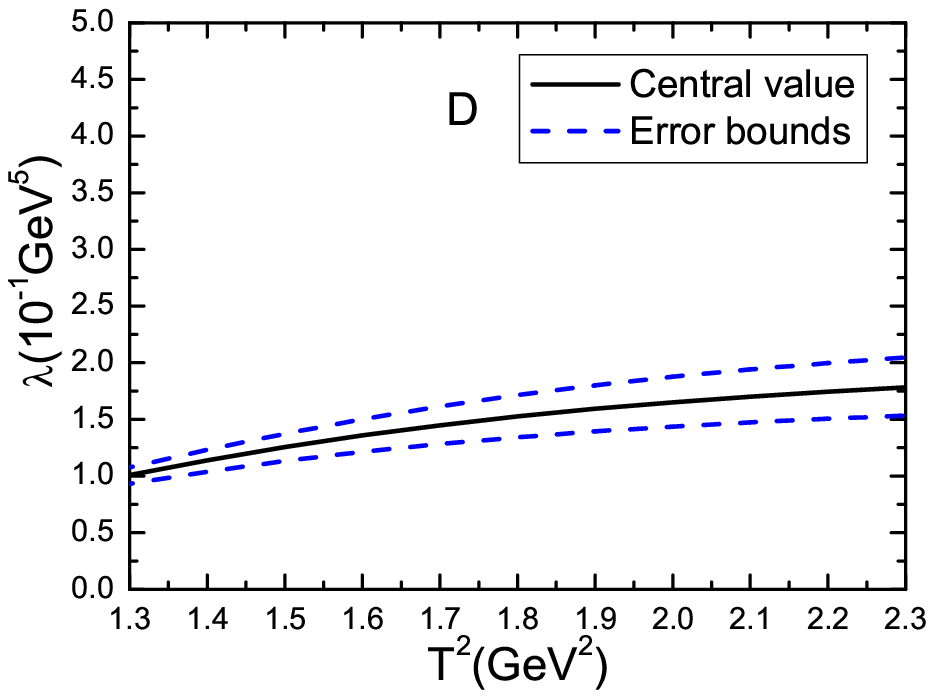}
        \caption{ The pole residues  of the charmed baryon states  with variations of the Borel parameters $T^2$, where the $A$, $B$, $C$ and $D$ correspond to the charmed baryon states $\Xi_c\left(1,1;\frac{5}{2}\right)$, $\Xi_c\left(1,1;\frac{3}{2}\right)$, $\Lambda_c\left(1,1;\frac{5}{2}\right)$ and $\Lambda_c\left(1,1;\frac{3}{2}\right)$, respectively.  }
\end{figure}

We take into account all uncertainties  of the input    parameters,
and obtain  the masses and pole residues of
 the  D-wave charmed baryon states $\Lambda_c$ and $\Xi_c$, which are shown explicitly in Figs.2-7 and
Table 2. In Figs.2-7,  we plot the masses and pole residues with variations
of the Borel parameters at much larger intervals   than the  Borel windows shown in Table 1. In the Borel windows, the uncertainties $\delta M/M$ originate from the Borel parameters are very small, about $(2-5)\%$, the Borel platforms exist approximately. Furthermore, the energy scale formula $ \mu =\sqrt{M_{B}^2-{\mathbb{M}}_c^2}$ is well satisfied.
The criteria $\bf{3}$ and $\bf{4}$ are satisfied, now the four criteria are all satisfied.

In Fig.2 and Table 2, we also present the experimental  values \cite{LHCb2860,PDG} and predictions from the QCD sum rules combined with the heavy quark effective theory \cite{Zhu-D-wave}.  The present predictions are consistent with the experimental values \cite{LHCb2860,PDG} and other QCD sum rules calculations \cite{Zhu-D-wave}, and support assigning the $\Lambda_c(2860)$, $\Lambda_c(2880)$, $\Xi_c(3055)$  and $\Xi_c(3080)$ to be the  D-wave charmed baryon states with the quantum numbers $(L_\rho,L_\lambda)=(0,2)$ and $J^P={\frac{3}{2}}^+$, ${\frac{5}{2}}^+$, ${\frac{3}{2}}^+$  and ${\frac{5}{2}}^+$, respectively. The predictions for the $(L_\rho,L_\lambda)=(2,0)$ and $(L_\rho,L_\lambda)=(1,1)$ D-wave $\Lambda_c$ and $\Xi_c$ states can be confronted to the experimental data in the future.

\section{Conclusion}
In this article, we tentatively assign the $\Lambda_c(2860)$, $\Lambda_c(2880)$, $\Xi_c(3055)$  and $\Xi_c(3080)$ to be the  D-wave charmed baryon states with  $J^P={\frac{3}{2}}^+$, ${\frac{5}{2}}^+$, ${\frac{3}{2}}^+$  and ${\frac{5}{2}}^+$, respectively, and study their masses and pole residues  with the QCD sum rules in a systematic way  by constructing three-types interpolating currents with the quantum numbers $(L_\rho,L_\lambda)=(0,2)$, $(2,0)$ and $(1,1)$, respectively. As the currents couple potentially to both the positive parity and negative parity baryon states, we separate the contributions of the ${\frac{3}{2}}^{\pm}$ and ${\frac{5}{2}}^{\pm}$ charmed baryon states unambiguously, and the QCD sum rules do not suffer from the contaminations of the charmed baryon states with  negative parity.  We carry out the operator product expansion up to  the vacuum condensates of dimension 10 in a consistent way, and use the  empirical energy scale formula to determine the optimal energy scales of the QCD spectral densities to extract the hadron masses.
The present predictions support assigning the $\Lambda_c(2860)$, $\Lambda_c(2880)$, $\Xi_c(3055)$  and $\Xi_c(3080)$ to be the  D-wave baryon states with the quantum numbers $(L_\rho,L_\lambda)=(0,2)$ and $J^P={\frac{3}{2}}^+$, ${\frac{5}{2}}^+$, ${\frac{3}{2}}^+$  and ${\frac{5}{2}}^+$, respectively. The predictions for the masses of the $(L_\rho,L_\lambda)=(2,0)$ and $(1,1)$ D-wave $\Lambda_c$ and $\Xi_c$ states can be confronted to the experimental data in the future.

\section*{Acknowledgements}
This  work is supported by National Natural Science Foundation, Grant Number 11375063.

\section*{Appendix}
The explicit expressions of the  QCD spectral densities $\rho^0_{j,QCD}(s)$ and $\rho^1_{j,QCD}(s)$,
\begin{eqnarray}
\rho^0_{j,QCD}(s)&=&m_c\,\rho^{0;\Xi_c/\Lambda_c}_{j,L_\rho,L_\lambda}(s)\, ,\nonumber\\
\rho^1_{j,QCD}(s)&=&\rho^{1;\Xi_c/\Lambda_c}_{j,L_\rho,L_\lambda}(s)\, ,
\end{eqnarray}

\begin{eqnarray}
\rho^{0;\Xi_c}_{{\frac{5}{2}},2,0}(s)&=&\frac{1}{69120\pi^4}\int_{x_i}^1dx \,(9x^2+34x+132)(1-x)^4(s-\widetilde{m}_c^2)^4 \nonumber\\
&&+\frac{5m_s\langle\bar{s}s\rangle-2m_s\langle\bar{q}q\rangle}{96\pi^2}\int_{x_i}^1dx \,x^2(1-x)^2(s-\widetilde{m}_c^2)^2 \nonumber\\
&&+\frac{m_s\langle\bar{q}g_s\sigma Gq\rangle}{36\pi^2}\int_{x_i}^1dx \,x(3x-4)(1-x)(s-\widetilde{m}_c^2) \nonumber\\
&&+\frac{m_s\langle\bar{s}g_s\sigma Gs\rangle}{216\pi^2}\int_{x_i}^1dx \,x(31-27x)(1-x)(s-\widetilde{m}_c^2) \nonumber\\
&&-\frac{m_c^2}{51840\pi^2}\langle\frac{\alpha_sGG}{\pi}\rangle\int_{x_i}^1dx \,\frac{(9x^2+34x+132)(1-x)^4}{x^3}(s-\widetilde{m}_c^2) \nonumber\\
&&+\frac{1}{34560\pi^2}\langle\frac{\alpha_sGG}{\pi}\rangle\int_{x_i}^1dx \,\frac{(9x^2+34x+132)(1-x)^4}{x^2}(s-\widetilde{m}_c^2)^2 \nonumber\\
&&+\frac{1}{6912\pi^2}\langle\frac{\alpha_sGG}{\pi}\rangle\int_{x_i}^1dx \,(9x^2+20x+46)(1-x)^2(s-\widetilde{m}_c^2)^2 \nonumber\\
&&+\frac{\langle\bar{q}g_s\sigma Gq\rangle\langle\bar{s}g_s\sigma Gs\rangle}{72}\delta(s-m_c^2) \, ,
\end{eqnarray}

\begin{eqnarray}
\rho^{1;\Xi_c}_{{\frac{5}{2}},2,0}(s)&=&\frac{1}{69120\pi^4}\int_{x_i}^1dx \,x(27x^2+55x+128)(1-x)^4(s-\widetilde{m}_c^2)^4 \nonumber\\
&&+\frac{5m_s\langle\bar{s}s\rangle-2m_s\langle\bar{q}q\rangle}{288\pi^2}\int_{x_i}^1dx \,x^2(9x-1)(1-x)^2(s-\widetilde{m}_c^2)^2 \nonumber\\
&&+\frac{m_s\langle\bar{q}g_s\sigma Gq\rangle}{72\pi^2}\int_{x_i}^1dx \,x(18x^2-17x+1)(1-x)(s-\widetilde{m}_c^2) \nonumber\\
&&+\frac{m_s\langle\bar{s}g_s\sigma Gs\rangle}{216\pi^2}\int_{x_i}^1dx \,x(-81x^2+70x-4)(1-x)(s-\widetilde{m}_c^2) \nonumber\\
&&-\frac{m_c^2}{51840\pi^2}\langle\frac{\alpha_sGG}{\pi}\rangle\int_{x_i}^1dx \,\frac{(27x^2+55x+128)(1-x)^4}{x^2}(s-\widetilde{m}_c^2) \nonumber\\
&&+\frac{1}{6912\pi^2}\langle\frac{\alpha_sGG}{\pi}\rangle\int_{x_i}^1dx \,x(27x^2+23x+40)(1-x)^2(s-\widetilde{m}_c^2)^2 \nonumber\\
&&+\frac{5\langle\bar{q}g_s\sigma Gq\rangle\langle\bar{s}g_s\sigma Gs\rangle}{432}\delta(s-m_c^2) \, ,
\end{eqnarray}

\begin{eqnarray}
\rho^{0;\Xi_c}_{{\frac{5}{2}},0,2}(s)&=&\frac{1}{4608\pi^4}\int_{x_i}^1dx \,x(3x-2)(1-x)^4(s-\widetilde{m}_c^2)^4 \nonumber\\
&&+\frac{m_s\langle\bar{s}s\rangle-2m_s\langle\bar{q}q\rangle}{96\pi^2}\int_{x_i}^1dx \,x^2(1-x)^2(s-\widetilde{m}_c^2)^2 \nonumber\\
&&+\frac{m_c^2}{3456\pi^2}\langle\frac{\alpha_sGG}{\pi}\rangle\int_{x_i}^1dx \,\frac{(2-3x)(1-x)^4}{x^2}(s-\widetilde{m}_c^2) \nonumber\\
&&+\frac{1}{2304\pi^2}\langle\frac{\alpha_sGG}{\pi}\rangle\int_{x_i}^1dx \,\frac{(3x-2)(1-x)^4}{x}(s-\widetilde{m}_c^2)^2 \nonumber\\
&&+\frac{1}{768\pi^2}\langle\frac{\alpha_sGG}{\pi}\rangle\int_{x_i}^1dx \,x^2(1-x)^2(s-\widetilde{m}_c^2)^2 \nonumber\\
&&+\frac{\langle\bar{q}g_s\sigma Gq\rangle\langle\bar{s}g_s\sigma Gs\rangle}{72}\delta(s-m_c^2) \, ,
\end{eqnarray}

\begin{eqnarray}
\rho^{1;\Xi_c}_{{\frac{5}{2}},0,2}(s)&=&\frac{1}{4608\pi^4}\int_{x_i}^1dx \,x^2(9x+1)(1-x)^4(s-\widetilde{m}_c^2)^4 \nonumber\\
&&+\frac{m_s\langle\bar{s}s\rangle-2m_s\langle\bar{q}q\rangle}{288\pi^2}\int_{x_i}^1dx \,x^2(9x-1)(1-x)^2(s-\widetilde{m}_c^2)^2 \nonumber\\
&&-\frac{m_c^2}{3456\pi^2}\langle\frac{\alpha_sGG}{\pi}\rangle\int_{x_i}^1dx \,\frac{(9x+1)(1-x)^4}{x}(s-\widetilde{m}_c^2) \nonumber\\
&&+\frac{1}{2304\pi^2}\langle\frac{\alpha_sGG}{\pi}\rangle\int_{x_i}^1dx \,x^2(9x-1)(1-x)^2(s-\widetilde{m}_c^2)^2 \nonumber\\
&&+\frac{5\langle\bar{q}g_s\sigma Gq\rangle\langle\bar{s}g_s\sigma Gs\rangle}{432}\delta(s-m_c^2) \, ,
\end{eqnarray}

\begin{eqnarray}
\rho^{0;\Xi_c}_{{\frac{5}{2}},1,1}(s)&=&\frac{1}{13824\pi^4}\int_{x_i}^1dx \,(3x^2+8x-3)(1-x)^4(s-\widetilde{m}_c^2)^4 \nonumber\\
&&+\frac{3m_s\langle\bar{s}s\rangle-2m_s\langle\bar{q}q\rangle}{96\pi^2}\int_{x_i}^1dx \,x^2(1-x)^2(s-\widetilde{m}_c^2)^2 \nonumber\\
&&+\frac{m_s\langle\bar{q}g_s\sigma Gq\rangle}{48\pi^2}\int_{x_i}^1dx \,x(2x-1)(1-x)(s-\widetilde{m}_c^2) \nonumber\\
&&-\frac{5m_s\langle\bar{s}g_s\sigma Gs\rangle}{432\pi^2}\int_{x_i}^1dx \,x(3x-1)(1-x)(s-\widetilde{m}_c^2) \nonumber\\
&&-\frac{m_c^2}{10368\pi^2}\langle\frac{\alpha_sGG}{\pi}\rangle\int_{x_i}^1dx \,\frac{(3x^2+8x-3)(1-x)^4}{x^3}(s-\widetilde{m}_c^2) \nonumber\\
&&+\frac{1}{6912\pi^2}\langle\frac{\alpha_sGG}{\pi}\rangle\int_{x_i}^1dx \,\frac{(3x^2+8x-3)(1-x)^4}{x^2}(s-\widetilde{m}_c^2)^2 \nonumber\\
&&+\frac{1}{4608\pi^2}\langle\frac{\alpha_sGG}{\pi}\rangle\int_{x_i}^1dx \,(6x^2+10x-1)(1-x)^2(s-\widetilde{m}_c^2)^2   \, ,
\end{eqnarray}

\begin{eqnarray}
\rho^{1;\Xi_c}_{{\frac{5}{2}},1,1}(s)&=&\frac{1}{13824\pi^4}\int_{x_i}^1dx \,x(9x^2+14x+2)(1-x)^4(s-\widetilde{m}_c^2)^4 \nonumber\\
&&+\frac{3m_s\langle\bar{s}s\rangle-2m_s\langle\bar{q}q\rangle}{288\pi^2}\int_{x_i}^1dx \,x^2(9x-1)(1-x)^2(s-\widetilde{m}_c^2)^2 \nonumber\\
&&+\frac{m_s\langle\bar{q}g_s\sigma Gq\rangle}{288\pi^2}\int_{x_i}^1dx \,x(36x^2-21x+1)(1-x)(s-\widetilde{m}_c^2) \nonumber\\
&&+\frac{m_s\langle\bar{s}g_s\sigma Gs\rangle}{432\pi^2}\int_{x_i}^1dx \,x(-45x^2+23x-1)(1-x)(s-\widetilde{m}_c^2) \nonumber\\
&&-\frac{m_c^2}{10368\pi^2}\langle\frac{\alpha_sGG}{\pi}\rangle\int_{x_i}^1dx \,\frac{(9x^2+14x+2)(1-x)^4}{x^2}(s-\widetilde{m}_c^2) \nonumber\\
&&+\frac{1}{4608\pi^2}\langle\frac{\alpha_sGG}{\pi}\rangle\int_{x_i}^1dx \,x^2(18x+11)(1-x)^2(s-\widetilde{m}_c^2)^2  \, ,
\end{eqnarray}

\begin{eqnarray}
\rho^{0;\Xi_c}_{{\frac{3}{2}},2,0}(s)&=&\frac{1}{3072\pi^4}\int_{x_i}^1dx \,(4x+33)(1-x)^4(s-\widetilde{m}_c^2)^4 \nonumber\\
&&+\frac{7m_s\langle\bar{s}g_s\sigma Gs\rangle-6m_s\langle\bar{q}g_s\sigma Gq\rangle}{24\pi^2}\int_{x_i}^1dx \,x(1-x)(s-\widetilde{m}_c^2) \nonumber\\
&&-\frac{m_c^2}{2304\pi^2}\langle\frac{\alpha_sGG}{\pi}\rangle\int_{x_i}^1dx \,\frac{(4x+33)(1-x)^4}{x^3}(s-\widetilde{m}_c^2) \nonumber\\
&&+\frac{1}{1536\pi^2}\langle\frac{\alpha_sGG}{\pi}\rangle\int_{x_i}^1dx \,\frac{(4x+33)(1-x)^4}{x^2}(s-\widetilde{m}_c^2)^2 \nonumber\\
&&+\frac{1}{768\pi^2}\langle\frac{\alpha_sGG}{\pi}\rangle\int_{x_i}^1dx \,(8x+31)(1-x)^2(s-\widetilde{m}_c^2)^2 \nonumber\\
&&+\frac{3\langle\bar{q}g_s\sigma Gq\rangle\langle\bar{s}g_s\sigma Gs\rangle}{32}\delta(s-m_c^2) \, ,
\end{eqnarray}

\begin{eqnarray}
\rho^{1;\Xi_c}_{{\frac{3}{2}},2,0}(s)&=&\frac{5}{3072\pi^4}\int_{x_i}^1dx \,x(4x+9)(1-x)^4(s-\widetilde{m}_c^2)^4 \nonumber\\
&&+\frac{25m_s\langle\bar{s}s\rangle-10m_s\langle\bar{q}q\rangle}{16\pi^2}\int_{x_i}^1dx \,x^2 (1-x)^2(s-\widetilde{m}_c^2)^2 \nonumber\\
&&+\frac{m_s\langle\bar{q}g_s\sigma Gq\rangle}{4\pi^2}\int_{x_i}^1dx \,x(7x-4)(1-x)(s-\widetilde{m}_c^2) \nonumber\\
&&+\frac{m_s\langle\bar{s}g_s\sigma Gs\rangle}{24\pi^2}\int_{x_i}^1dx \,x(33-64x)(1-x)(s-\widetilde{m}_c^2) \nonumber\\
&&-\frac{5m_c^2}{2304\pi^2}\langle\frac{\alpha_sGG}{\pi}\rangle\int_{x_i}^1dx \,\frac{(4x+9)(1-x)^4}{x^2}(s-\widetilde{m}_c^2) \nonumber\\
&&+\frac{1}{768\pi^2}\langle\frac{\alpha_sGG}{\pi}\rangle\int_{x_i}^1dx \,x(34x+35)(1-x)^2(s-\widetilde{m}_c^2)^2 \nonumber\\
&&+\frac{5\langle\bar{q}g_s\sigma Gq\rangle\langle\bar{s}g_s\sigma Gs\rangle}{96}\delta(s-m_c^2) \, ,
\end{eqnarray}

\begin{eqnarray}
\rho^{0;\Xi_c}_{{\frac{3}{2}},0,2}(s)&=&\frac{1}{1024\pi^4}\int_{x_i}^1dx \,(3-4x)(1-x)^4(s-\widetilde{m}_c^2)^4 \nonumber\\
&&-\frac{m_c^2}{768\pi^2}\langle\frac{\alpha_sGG}{\pi}\rangle\int_{x_i}^1dx \,\frac{(3-4x)(1-x)^4}{x^3}(s-\widetilde{m}_c^2) \nonumber\\
&&+\frac{1}{512\pi^2}\langle\frac{\alpha_sGG}{\pi}\rangle\int_{x_i}^1dx \,\frac{(3-4x)(1-x)^4}{x^2}(s-\widetilde{m}_c^2)^2 \nonumber\\
&&+\frac{3\langle\bar{q}g_s\sigma Gq\rangle\langle\bar{s}g_s\sigma Gs\rangle}{32}\delta(s-m_c^2) \, ,
\end{eqnarray}

\begin{eqnarray}
\rho^{1;\Xi_c}_{{\frac{3}{2}},0,2}(s)&=&\frac{7}{1024\pi^4}\int_{x_i}^1dx \,x(4x+1)(1-x)^4(s-\widetilde{m}_c^2)^4 \nonumber\\
&&+\frac{5m_s\langle\bar{s}s\rangle-10m_s\langle\bar{q}q\rangle}{16\pi^2}\int_{x_i}^1dx \,x^2 (1-x)^2(s-\widetilde{m}_c^2)^2 \nonumber\\
&&-\frac{7m_c^2}{768\pi^2}\langle\frac{\alpha_sGG}{\pi}\rangle\int_{x_i}^1dx \,\frac{(4x+1)(1-x)^4}{x^2}(s-\widetilde{m}_c^2) \nonumber\\
&&+\frac{5}{128\pi^2}\langle\frac{\alpha_sGG}{\pi}\rangle\int_{x_i}^1dx \,x^2(1-x)^2(s-\widetilde{m}_c^2)^2 \nonumber\\
&&+\frac{5\langle\bar{q}g_s\sigma Gq\rangle\langle\bar{s}g_s\sigma Gs\rangle}{96}\delta(s-m_c^2) \, ,
\end{eqnarray}

\begin{eqnarray}
\rho^{0;\Xi_c}_{{\frac{3}{2}},1,1}(s)&=&\frac{1}{768\pi^4}\int_{x_i}^1dx \,(x-3)(1-x)^4(s-\widetilde{m}_c^2)^4 \nonumber\\
&&-\frac{m_s\langle\bar{s}g_s\sigma Gs\rangle}{48\pi^2}\int_{x_i}^1dx \,x(1-x)(s-\widetilde{m}_c^2) \nonumber\\
&&-\frac{m_c^2}{576\pi^2}\langle\frac{\alpha_sGG}{\pi}\rangle\int_{x_i}^1dx \,\frac{(x-3)(1-x)^4}{x^3}(s-\widetilde{m}_c^2) \nonumber\\
&&+\frac{1}{384\pi^2}\langle\frac{\alpha_sGG}{\pi}\rangle\int_{x_i}^1dx \,\frac{(x-3)(1-x)^4}{x^2}(s-\widetilde{m}_c^2)^2 \nonumber\\
&&+\frac{1}{1024\pi^2}\langle\frac{\alpha_sGG}{\pi}\rangle\int_{x_i}^1dx \,(8x-5)(1-x)^2(s-\widetilde{m}_c^2)^2   \, ,
\end{eqnarray}

\begin{eqnarray}
\rho^{1;\Xi_c}_{{\frac{3}{2}},1,1}(s)&=&\frac{1}{768\pi^4}\int_{x_i}^1dx \,x(8x+7)(1-x)^4(s-\widetilde{m}_c^2)^4 \nonumber\\
&&+\frac{15m_s\langle\bar{s}s\rangle-10m_s\langle\bar{q}q\rangle}{16\pi^2}\int_{x_i}^1dx \,x^2 (1-x)^2(s-\widetilde{m}_c^2)^2 \nonumber\\
&&+\frac{5m_s\langle\bar{q}g_s\sigma Gq\rangle}{16\pi^2}\int_{x_i}^1dx \,x(3x-1)(1-x)(s-\widetilde{m}_c^2) \nonumber\\
&&+\frac{m_s\langle\bar{s}g_s\sigma Gs\rangle}{48\pi^2}\int_{x_i}^1dx \,x(11-38x)(1-x)(s-\widetilde{m}_c^2) \nonumber\\
&&-\frac{m_c^2}{576\pi^2}\langle\frac{\alpha_sGG}{\pi}\rangle\int_{x_i}^1dx \,\frac{(8x+7)(1-x)^4}{x^2}(s-\widetilde{m}_c^2) \nonumber\\
&&+\frac{1}{1024\pi^2}\langle\frac{\alpha_sGG}{\pi}\rangle\int_{x_i}^1dx \,x(44x+15)(1-x)^2(s-\widetilde{m}_c^2)^2  \, ,
\end{eqnarray}

\begin{eqnarray}
\rho^{0;\Lambda_c}_{{\frac{5}{2}},2,0}(s)&=& \rho^{0;\Xi_c}_{{\frac{5}{2}},2,0}(s)\mid_{m_s\to 0, \,\, \langle\bar{s}s\rangle \to \langle\bar{q}q\rangle, \,\,\langle\bar{s}g_s\sigma Gs\rangle \to \langle\bar{q}g_s\sigma Gq\rangle } \, , \nonumber\\
\rho^{0;\Lambda_c}_{{\frac{5}{2}},0,2}(s)&=& \rho^{0;\Xi_c}_{{\frac{5}{2}},0,2}(s)\mid_{m_s\to 0, \,\, \langle\bar{s}s\rangle \to \langle\bar{q}q\rangle, \,\,\langle\bar{s}g_s\sigma Gs\rangle \to \langle\bar{q}g_s\sigma Gq\rangle } \, , \nonumber\\
\rho^{0;\Lambda_c}_{{\frac{5}{2}},1,1}(s)&=& \rho^{0;\Xi_c}_{{\frac{5}{2}},1,1}(s)\mid_{m_s\to 0, \,\, \langle\bar{s}s\rangle \to \langle\bar{q}q\rangle, \,\,\langle\bar{s}g_s\sigma Gs\rangle \to \langle\bar{q}g_s\sigma Gq\rangle } \, , \nonumber\\
\rho^{1;\Lambda_c}_{{\frac{5}{2}},2,0}(s)&=& \rho^{1;\Xi_c}_{{\frac{5}{2}},2,0}(s)\mid_{m_s\to 0, \,\, \langle\bar{s}s\rangle \to \langle\bar{q}q\rangle, \,\,\langle\bar{s}g_s\sigma Gs\rangle \to \langle\bar{q}g_s\sigma Gq\rangle } \, , \nonumber\\
\rho^{1;\Lambda_c}_{{\frac{5}{2}},0,2}(s)&=& \rho^{1;\Xi_c}_{{\frac{5}{2}},0,2}(s)\mid_{m_s\to 0, \,\, \langle\bar{s}s\rangle \to \langle\bar{q}q\rangle, \,\,\langle\bar{s}g_s\sigma Gs\rangle \to \langle\bar{q}g_s\sigma Gq\rangle } \, , \nonumber\\
\rho^{1;\Lambda_c}_{{\frac{5}{2}},1,1}(s)&=& \rho^{1;\Xi_c}_{{\frac{5}{2}},1,1}(s)\mid_{m_s\to 0, \,\, \langle\bar{s}s\rangle \to \langle\bar{q}q\rangle, \,\,\langle\bar{s}g_s\sigma Gs\rangle \to \langle\bar{q}g_s\sigma Gq\rangle } \, ,
\end{eqnarray}

\begin{eqnarray}
\rho^{0;\Lambda_c}_{{\frac{3}{2}},2,0}(s)&=& \rho^{0;\Xi_c}_{{\frac{3}{2}},2,0}(s)\mid_{m_s\to 0, \,\, \langle\bar{s}s\rangle \to \langle\bar{q}q\rangle, \,\,\langle\bar{s}g_s\sigma Gs\rangle \to \langle\bar{q}g_s\sigma Gq\rangle } \, , \nonumber\\
\rho^{0;\Lambda_c}_{{\frac{3}{2}},0,2}(s)&=& \rho^{0;\Xi_c}_{{\frac{3}{2}},0,2}(s)\mid_{m_s\to 0, \,\, \langle\bar{s}s\rangle \to \langle\bar{q}q\rangle, \,\,\langle\bar{s}g_s\sigma Gs\rangle \to \langle\bar{q}g_s\sigma Gq\rangle } \, , \nonumber\\
\rho^{0;\Lambda_c}_{{\frac{3}{2}},1,1}(s)&=& \rho^{0;\Xi_c}_{{\frac{3}{2}},1,1}(s)\mid_{m_s\to 0, \,\, \langle\bar{s}s\rangle \to \langle\bar{q}q\rangle, \,\,\langle\bar{s}g_s\sigma Gs\rangle \to \langle\bar{q}g_s\sigma Gq\rangle } \, , \nonumber\\
\rho^{1;\Lambda_c}_{{\frac{3}{2}},2,0}(s)&=& \rho^{1;\Xi_c}_{{\frac{3}{2}},2,0}(s)\mid_{m_s\to 0, \,\, \langle\bar{s}s\rangle \to \langle\bar{q}q\rangle, \,\,\langle\bar{s}g_s\sigma Gs\rangle \to \langle\bar{q}g_s\sigma Gq\rangle } \, , \nonumber\\
\rho^{1;\Lambda_c}_{{\frac{3}{2}},0,2}(s)&=& \rho^{1;\Xi_c}_{{\frac{3}{2}},0,2}(s)\mid_{m_s\to 0, \,\, \langle\bar{s}s\rangle \to \langle\bar{q}q\rangle, \,\,\langle\bar{s}g_s\sigma Gs\rangle \to \langle\bar{q}g_s\sigma Gq\rangle } \, , \nonumber\\
\rho^{1;\Lambda_c}_{{\frac{3}{2}},1,1}(s)&=& \rho^{1;\Xi_c}_{{\frac{3}{2}},1,1}(s)\mid_{m_s\to 0, \,\, \langle\bar{s}s\rangle \to \langle\bar{q}q\rangle, \,\,\langle\bar{s}g_s\sigma Gs\rangle \to \langle\bar{q}g_s\sigma Gq\rangle } \, ,
\end{eqnarray}
$\widetilde{m}_c^2=\frac{m_c^2}{x}$, $x_i=\frac{m_c^2}{s}$.

\end{document}